\documentstyle[12pt]{article}
\input{epsf}
\begin{document}
\title{Effective action of quantum fields in the space-time of a cylindrically
symmetric spinning body}
\author{J.A. Briginshaw\thanks{%
e-mail jab24@damtp.cam.ac.uk} \\
%EndAName
{}\\Department of Applied Mathematics and \\ Theoretical Physics\\
Silver Street, Cambridge, CB3 9EW, UK}
\maketitle
\begin{abstract}
The aim of this article is to calculate (to first order in $\hbar$)
the renormalized effective action of a self interacting massive scalar
field propagating in the space-time due to a cylindrically symmetric,
rotating body. The vacuum (exterior space-time) contribution is model
independent, we also consider the simplest case of a core (interior 
space-time) model, namely a cylindrical shell. The heat kernel of the 
system is calculated, and used to obtain an expression for the
determinant of the Klein-Gordon operator on the space-time manifold. 
New ultra-violet poles are discovered, and regularization techniques 
are then employed to render finite the Klein-Gordon determinant and 
consequently extract the regularized one loop effective action for a 
self interacting scalar field theory. The coupling constants of the 
theory are then renormalized. As a test case a conical singularity
with non-zero flux is also considered.
\end{abstract}


\pagebreak

\section{Introduction}

\paragraph{}
The use of manifolds containing conical singularities to model (infinitely)
thin line sources has now become quite common. For example, a space-time
metric which is flat except for a single conical singularity is a reasonable
starting point in the investigation of the physics of cosmic strings -
topological defects which may be generated by phase transitions occurring in
certain grand unified models \cite{kib1}. Conical singularities also arise
when considering the thermodynamic properties of black holes. By making the
integral curves of the (previously time-like) Killing vector in a 
Euclideanized static black hole metric suitably periodic, it is possible to 
construct a partition function for a quantum field propagating on the
black hole background. The contribution of the quantum field to the
black hole entropy can then be easily obtained by standard
thermodynamic methods \cite{bir1} \cite{man1} \cite{wal1}. In addition
to these applications in four dimensions, it has been recognized that 
conical singularities play a crucial role in 2+1 dimensional gravity 
\cite{des2} \cite{ger2}. The gravitational field of the Einstein
theory in 2+1 dimensions has no dynamical degrees of freedom
(i.e. there is no graviton in the theory) and gravitational
interaction is possible only by the fact that every massive particle
has an associated conical singularity. The more massive
the particle, the more ``pointed'' the cone it induces on the
space-time manifold. Two particle scattering is then facilitated simply by 
each particle travelling on the cone generated by the other.

\paragraph{}
With regard to the conical singularity as a model of cosmic strings, one of
the less pleasing features of the infinitely thin idealization is that there
is a delta function singularity of Riemannian curvature at the position
of the string itself (see \cite{ger2} or section 12 of this paper). Whilst
this is not necessarily fatal, indeed there are now well defined procedures
for dealing with this issue, it is more appealing to regard the conical
metric as being the exterior space-time to some finite sized cylindrical
body, which is suitably well behaved at the origin of coordinates. Naturally
appropriate boundary conditions are required to patch together the interior
and exterior solutions (\cite{jen2} or section 12) and the result is a
metric describing the space-time of a finitely thick but infinitely
long cylindrical body.

\paragraph{}
Another natural generalization is the space-time of a rotating,
cylindrically symmetric body. For example, it has already been
shown that in a particular model of a superconducting cosmic string, the
angular momentum of the left moving and right moving modes may not
cancel, resulting in a string with net angular momentum \cite{maz2}. Another
example occurs in Chern-Simons field theory, where string-like objects with
angular momentum can also occur \cite{cho1}. Additionally it is
hard to imagine a mechanism which would prevent one simply adding angular
momentum to a finite thickness cylindrical body by hand, although naturally
one must bear the infinite length of such an object in mind.

\paragraph{}
The natural starting place for this problem would seem to be to
generalize the metric describing a simple conical singularity by adding
angular momentum to the space-time. This was duly done \cite{ger1}, but threw 
up some surprising results. The constructed metric contained closed
time-like curves concentrated in a finite region around the axis of
symmetry, and it was soon realized that in addition to the curvature 
singularity, this model had a torsion singularity on axis as well
\cite{ger2} \cite{tod1}. From then on approaches diverged as to how
best to treat the problem. With the appearance of torsion, some
authors argued that this metric was best considered within the
framework of Einstein-Cartan theory, which naturally incorporates
torsion as well as curvature \cite{sol3}. Models were constructed within
Einstein-Cartan theory to describe the space-time of a conical singularity
with spin \cite{jen2}, whilst other authors chose a different direction.
Discovering that the first quantized scattering problem on the manifold was
badly defined within the closed time-like curve region \cite{ger1}, it was
argued that by restricting the angular momentum of the string to certain
values (and hence invoking a quantization condition), it is possible to make
the string ``transparent'' to scattering experiments, and hence remove the
difficulty with the closed time-like curve region \cite{maz1}. Others argued
for the presence of angular momentum dependence in the first quantum
correction to the vacuum energy, with analogy to the case of the spinning
circle \cite{jen1}, or simply on dimensional grounds \cite{mat1}.

\paragraph{}
In this paper we will take a different point of view. It is clearly the
region immediately around the axis which causes the difficulties. The
Hamiltonian of scalar quantum fields is non-Hermitian here, both in quantum
mechanics \cite{ger1} and in quantum field theory (see section 11) . More
convincingly, the results of Menotti and Seminara (\cite{men1} and
following papers) indicate that if the energy-momentum tensor of the
space-time satisfies the weak energy condition, and if we have no closed
time-like curves (CTC's) at spatial infinity (which must be conical and open
in nature), then there are no CTC's in the space-time at all. In other
words, any energy-momentum tensor which we write down as a source of the
spinning conical space-time will be non-physical (violate the weak energy
condition) if we include the region of closed time-like curves. 
Thus we will consider the spinning conical space-time only as
the exterior solution to some interior space-time, which extends
beyond the closed time-like curve region and is torsion free.

\paragraph{}
There are a variety of models that one can choose for the interior
space-time (\cite{jen2}, but note \cite{sol2}). Since our eventual aim is
to calculate the effective action for a massive scalar quantum field
propagating on the manifold, the interior contribution to the effective
action will be model dependent. Therefore we will initially concentrate on
the exterior (model independent) terms, returning to consider the interior
metric again in section 12, where we will look at the particular case of the
cylindrical shell.

\section{\bf The metric of the exterior space-time}

\paragraph{}
The metric describing a simple conical singularity at the origin of
co-ordinates can be written in several ways (see, for example, \cite{des1}).
The representation which will be used in this paper can be expressed
in cylindrical polar co-ordinates as follows,

\begin{equation}
ds^2 = -dt^2 + d \rho^2 + \rho^2 d \phi^2 + d z^2 \hspace{1cm} ,
\end{equation}

\noindent where $(t,\rho ,\phi ,z)=(t,\rho ,\phi +2\pi \alpha ,z)$, $0<\rho <\infty
 $, $0<\phi <2\pi \alpha $, $-\infty <t<\infty $, $-\infty <z<\infty $
 and $0<\alpha <1$. The form of this metric makes it obvious that the
 space-time is locally flat, but it differs globally from Minkowski space
 (expressed in cylindrical polar co-ordinates) because of the
 identification in the angular variable, $\phi$. This condition
 means that a ``wedge'' (or deficit angle) is removed from the range of 
 $\phi$, and then the edges of the excised region are
 identified, resulting in a conical manifold. Clearly $\alpha $ is
 related to the magnitude of the deficit angle, and it dictates the
 degree of ``sharpness'' of the cone. If $\alpha=1$ then there is no
 deficit angle and we recover the metric for Minkowski space.

\paragraph{}
In the above space-time the source is envisaged to be an infinitely long,
infinitesimally thin tube along the $z$-axis of the co-ordinate set. Whilst
the space-time is locally flat away from the origin, there is in
fact a delta function singularity of curvature at the origin
(see \cite{ger2} or section 12). With regard to the physical
parameters relating to the source, $\alpha $ may be expressed in terms
of the mass per unit length $ \mu $ by the relation 
$ \alpha = 1-4G \mu /c^2 $. From now on factors of $c$, $\hbar $, $G$
and so on will be suppressed in all expressions except where noted.

\paragraph{}
In order to consider rotating sources, metric (1) needs to be 
generalized to include an angular momentum parameter. This parameter
describes the effect that source spin has on the exterior space-time.
The metric for a conical space-time with a spinning source takes the 
form

\begin{eqnarray}
ds^2 &=&-(dt+\frac{S}\alpha d\phi )^2+d\rho ^{\prime }{}^2+(\rho
^{\prime }+k)^2d\phi ^2+dz^{\prime}{}^2  \nonumber \\
&=&-dt^2-\frac{2S}\alpha dtd\phi +d{\rho }^{\prime }{}^2+\left( (\rho
^{\prime }+k)^2-\frac{S^2}{\alpha ^2}\right) d\phi ^2+dz^{\prime}{}^2 
\hspace{0.5cm},
\end{eqnarray}

\noindent where $S$ parameterizes the angular momentum and has the
dimensions of length \cite{des2}, $k$ is a constant \cite{jen2}, $(t,\rho ^{\prime },\phi ,z^{\prime})=(t,\rho ^{\prime },\phi
+2\pi \alpha ,z^{\prime})$ , $-k<\rho ^{\prime }<\infty $, $0<\phi <2\pi \alpha $%
, $-\infty <t<\infty $, $-\infty <z^{\prime}<\infty $ and $0<\alpha
<1$. The precise relationship between $S$ and the physical angular
momentum per unit length, $J$, can be written as $S=4GJ/c^3$.

\paragraph{}
In addition to the curvature singularity, it can be shown that this metric
also contains a torsion delta function singularity at the origin 
($\rho ^{\prime }=-k$) \cite{ger2} \cite{tod1}.

\paragraph{}
The space-time described by metric (2) clearly possesses closed time-like
curves (for example the contour described by $dt=0$, $d\rho ^{\prime }=0$, $%
dz^{\prime}=0$, $d\phi $ = constant, $(\rho ^{\prime }+k)<\frac{|S|}\alpha $
is closed and time-like). Whilst it has been shown that classically there
are no closed time-like geodesics in this space-time \cite{des1}, in the
path integral formulation of field theory one is supposed to include
weighted contributions from all paths, including those non classical paths
described by the closed time-like curves. We would therefore expect
unitarity problems describing physics on this manifold. As a consequence we
restrict the radial co-ordinate $\rho ^{\prime }>\rho _0>\frac{|S|}%
\alpha -k$. This is sufficient to remove the difficulties with the closed
time-like curves and will allow us to discuss quantum field theory on the
manifold. The rest of the space-time ($\rho ^{\prime }<\rho _0$) is supposed
to be composed of the source core, which is described by another metric and
must be matched to this one with appropriate conditions at the boundary (%
\cite{jen2} or section 12). The included constant $k$ is determined by
correctly matching the components of the two metrics at the boundary $\rho
^{\prime }=\rho _0$.

\paragraph{}
It should be noted that this metric is stationary, but not static. The
applicability of conical methods to stationary but non static space-times
has not yet been proved in general, but there is some strong evidence to
indicate that these methods hold there too. For example, the first quantum
correction to the effective action on the Kerr-Newman black hole background
has been obtained using conical methods \cite{man1}, after verifying that
the induced conical singularity still behaved in a distributional sense.
Since it has already been demonstrated that the conical singularity in this
space-time behaves in a distribution manner (\cite{ger2} \cite{tod1} and
section 12) we can have confidence that application of conical methods will
yield useful results.

\paragraph{}
We can define a new time co-ordinate $\overline{t}$ such that
$d\overline{t}=dt+ \frac{S}\alpha d\phi $ (or alternatively
$\overline{t}=t+\frac{S}{\alpha} \phi $). Additionally we can 
write $\rho^{\prime \prime} =\rho ^{\prime }+k$. With
these co-ordinates we can re-write metric (2) as

\begin{equation}
ds^2 = -d \overline{t}^2 + d \rho^{\prime \prime}{}^2 +\rho^{\prime
  \prime}{}^2d \phi^2 + d z^{\prime}{}^2 \hspace{1cm}, 
\end{equation}

\noindent where we make the following identifications,
  $(\overline{t},\rho^{\prime \prime},\phi ,z)=( \overline{t}+2\pi S,
\rho^{\prime \prime} ,\phi +2\pi \alpha ,z)$, $\rho _0+k<\rho^{\prime \prime}
<\infty $, $0<\phi <2\pi \alpha $, $-\infty <\overline{t}<\infty $, $%
-\infty < z^{\prime} < \infty $ and $ \rho _0>\frac{|S|}\alpha -k $.

\paragraph{}
The manifold as we have defined it above excludes the origin and the closed
time-like curve region. Therefore it has no curvature, no torsion, and
quantum theory on this manifold should avoid potential unitarity problems.

\paragraph{}
It will be convenient in what follows to perform the calculations in a
space-time with a Euclidean signature. By continuing the 
$\rho^{\prime \prime}$, $z^{\prime} $ co-ordinates to imaginary values, 
we may rewrite the metric as follows,

\begin{equation}
ds^2 = -d \overline{t}^2 - d \rho^2 - \rho^2 d \phi^2 - d z^2 \hspace{1cm},
\end{equation}

\noindent where $(\overline{t},\rho ,\phi ,z)=(\overline{t}+2\pi S,\rho ,\phi +2\pi
\alpha ,z)$, $\overline{\rho} _0+\overline{k}<\rho <\infty $, $0<\phi <2\pi \alpha $, $%
-\infty <\overline{t}<\infty $, $-\infty <z<\infty $ and $
|\overline{\rho} _0| > \frac{|S|}\alpha -k $. 

\paragraph{}
The parameters are
related to each other by $ \rho^{\prime \prime} =i \rho $, 
$ z^{\prime}=iz $, $\overline{k} =ik $, $\overline{\rho} _0 = i \rho_0 $.

\section{\bf Green's function and heat kernel}

\paragraph{}

We define the differential operator $\Box _x$ on our manifold by the expression

\begin{equation}
\Box_x = -\frac{\partial^2}{{\partial \overline{t}}^2} - \frac{\partial^2}{{%
\partial z}^2} - \frac{1}{\rho} \frac{\partial}{\partial \rho} \left( \rho 
\frac{\partial}{\partial \rho} \right) - \frac{1}{\rho^2} \frac{ \partial^2}{%
{\partial \phi}^2} \hspace{1cm},
\end{equation}

\noindent where $x=(\overline{t},\rho ,\phi ,z)$, and we define the
Euclidean Feynman Propagator $G_F^E(x,x^{\prime })$ for a scalar field
of mass $M$ by the equation

\begin{equation}
\left( \Box_x + M^2 \right) G_F^E (x,x^{\prime}) = -\delta^{(4)}
(x-x^{\prime}) \hspace{1cm}.
\end{equation}

\noindent We may regard $\Box _x+M^2$ as the Euclidean Klein-Gordon operator on the
manifold, the four dimensional delta function can be expressed as follows,

\begin{equation}
\delta^{(4)} (x-x^{\prime}) = \frac{1}{\rho} \delta \left( \overline{t} - 
\overline{t^{\prime}} - S \frac{(\phi - \phi^{\prime})}{\alpha} \right)
\delta (\rho - \rho^{\prime}) \delta (\phi - \phi^{\prime}) \delta
(z-z^{\prime}) \hspace{0.5cm}.
\end{equation}

\paragraph{}
The eigenfunctions of the operator $\Box _x+M^2$ with the correct
cylindrical symmetry and angular behaviour may be written as follows, \cite{mor1}

\begin{equation}
\Psi_{\omega, k, k_3, m} (x)= \frac{1}{(2 \pi)^{\frac{3}{2}} \alpha^{\frac{1%
}{2}} } J_{\frac{|m+\omega S|}{\alpha}} (k \rho) e^{i \left( k_3 z - \omega 
\overline{t} + (m+\omega S) \frac{\phi}{\alpha} \right)} \hspace{1cm} ,
\end{equation}

\noindent with eigenvalues

\begin{equation}
E_{\omega, k, k_3, m} = \omega^2 +k_3^2 +k^2 +M^2 \hspace{1cm} .
\end{equation}

\noindent $J_\nu (z)$ is a Bessel function of the first kind. Many of
the properties of the special functions used in this paper can be
located either in \cite {aba1} or \cite{gra1}. By standard
Sturm-Liouville theory we may immediately write down the Euclidean
Feynman propagator for massive scalar fields on this space-time as

\begin{eqnarray}
G_F^E(x,x^{\prime }) &=& -\int_0^\infty d\tau \sum_{m=-\infty }^\infty
\int_{-\infty }^\infty d\omega \int_{-\infty }^\infty dk_3\int_0^\infty kdk 
\nonumber \\
&& \hspace{2cm} e^{-\tau E_{\omega ,k,k_3,m}}\Psi _{\omega ,k,k_3,m}(x)\Psi _{\omega
,k,k_3,m}^{*}(x^{\prime }) \hspace{0.5cm} ,
\end{eqnarray}

\noindent where $\Psi ^{*}$ denotes the complex conjugate of $\Psi $
and $\tau $ takes the role of a Schwinger parameter.

\paragraph{}
There is a relationship between the Euclidean heat kernel $F^E(x,x^{\prime
},\tau )$ for the diffusion problem on this manifold and the Euclidean
Green's function, namely

\begin{equation}
G_F^{E} (x, x^{\prime}) = - \int_0^{\infty} d \tau F^E (x, x^{\prime}, \tau )
\hspace{1cm} . 
\end{equation}

\noindent If we note the following integral identities,

\begin{equation}
\int_{-\infty}^{\infty} d k_3 e^{- \tau k_3^2 + i k_3 (z-z^{\prime})} = 
\sqrt{\frac{\pi}{\tau}} e^{-\frac{(z-z^{\prime})^2}{4 \tau} }
\hspace{1cm} ,
\end{equation}

\begin{equation}
\int_0^\infty kdkJ_{\frac{|m+\omega S|}\alpha }(k\rho )J_{\frac{|m+\omega S|}%
\alpha }(k\rho ^{\prime })e^{-\tau k^2}=\frac 1{2\tau }e^{-\frac{(\rho
^2+\rho ^{\prime }{}^2)}{4\tau }}I_{\frac{|m+\omega S|}\alpha }\left( \frac{%
\rho \rho ^{\prime }}{2\tau }\right) \hspace{0.5cm},
\end{equation}

\noindent where $I_\nu (z)$ is a modified Bessel function of the first
kind, we may write an expression for the Euclidean heat kernel of this problem as follows,

\begin{eqnarray}
F^E(x,x^{\prime },\tau ) = \frac 1{(2\pi )^3\alpha (2\tau )}\sqrt{\frac \pi
\tau }e^{-\frac{(\rho ^2+\rho ^{\prime }{}^2)}{4\tau }-\frac{(z-z^{\prime
})^2}{4\tau }-M^2\tau }  \hspace{2.5cm} \nonumber \\ \hspace{1cm}
\int_{-\infty }^\infty d\omega e^{-\tau \omega ^2-i\omega (\overline{t}-
\overline{t}^{\prime }-\frac S\alpha (\phi -\phi ^{\prime
}))}\sum_{m=-\infty }^\infty I_{\frac{|m+\omega S|}\alpha }\left( \frac{\rho
\rho ^{\prime }}{2\tau }\right) e^{im\frac{(\phi -\phi ^{\prime
  })}\alpha }. 
\end{eqnarray}

\section{\bf Simplification of the heat kernel}

\paragraph{}
Having found an expression for the Euclidean heat kernel of the system, our
intention is to use it to find a regularized expression for the determinant
of the Klein-Gordon operator $ \Box _x+M^2 $ on this manifold. This in turn
will allow us to write down the first quantum correction term in the effective
action (see section 11).

\paragraph{}
Before we can proceed with this operation, we need to find a simpler
expression for the heat kernel, which can be more easily manipulated. In
order to proceed further we must invoke a complex representation for the
modified Bessel function $I_\nu (x)$,

\begin{equation}
I_{\frac{|m+\omega S|}{\alpha}} \left(\frac{\rho \rho^{\prime}}{2 \tau}
\right) = \frac{1}{2 \pi} \int_{C_1} e^{iz \left(\frac{|m+\omega S|}{\alpha}%
\right) +\frac{\rho \rho^{\prime}}{2 \tau} \cos z } dz \hspace{1cm},
\end{equation}

\noindent where $C_1$ is the contour depicted above in Figure 1.

\begin{figure}
\epsffile{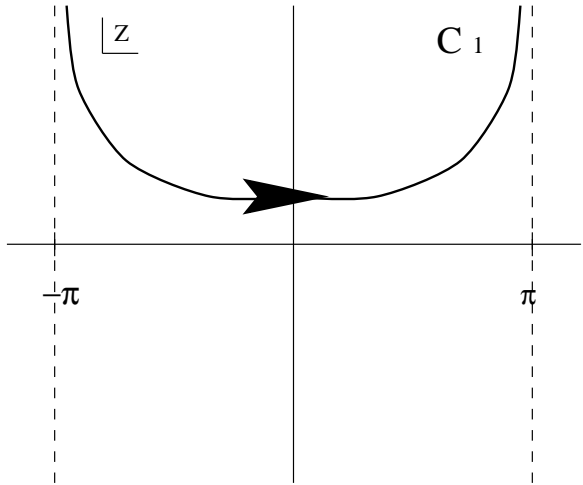}
\caption{Integration contour $C_1$ in the complex z plane}
\end{figure}

\paragraph{}
We now gather together terms in the expression for the heat kernel which
depend on $m$ and perform the summation over $m$, i.e. we
evaluate the following sum,

\begin{equation}
\sum_{m=-\infty }^\infty e^{iz\frac{|m+\omega S|}\alpha +im\frac{(\phi -\phi
^{\prime })}\alpha } \hspace{1cm}.
\end{equation}

\paragraph{}
Let us define the following functions, $ {\rm Int}(X)$ and ${\rm
Frac}(X)$ for real $X$. They are defined such that $X={\rm
Int}(X)+{\rm Frac} (X)$, ${\rm Int} (X)$ gives the
integer part of $X$ and ${\rm Frac}(X)$ gives the fractional part of $X$, 
where $0<{\rm Frac}(X)<1$ for both positive and negative $X$. So for example
${\rm Frac}(-5.2)=0.8$ and ${\rm Int}(-5.2)=-6$. It is then clear that the above
sum can be re-written (using $m \mapsto m-{\rm Int}(\omega S)$) as

\begin{equation}
e^{-i {\rm Int} (\omega S) \frac{(\phi-\phi^{\prime})}{\alpha}}
\sum_{m=-\infty}^{\infty} e^{iz \frac{|m+{\rm Frac}(\omega S)|}{\alpha} +im \frac{%
(\phi-\phi^{\prime})}{\alpha} } \hspace{1cm} .
\end{equation}

\noindent This sum can be evaluated by standard means, and gives the result

\begin{equation}
e^{-i {\rm Int} (\omega S) \frac{(\phi-\phi^{\prime})}{\alpha} } \left( \frac{
e^{iz \frac{{\rm Frac}(\omega S)}{\alpha} } }{1-e^{\frac{i}{\alpha}
(z+(\phi-\phi^{\prime}))}} + \frac{e^{-iz \frac{{\rm Frac}(\omega S)}{\alpha} } }{%
e^{-\frac{i}{\alpha} (z- (\phi-\phi^{\prime}))}-1} \right) \hspace{1cm}.
\end{equation}

\paragraph{}
The two terms resulting from the above summation are very similar, and
if we let $z \mapsto -z$ in one of the terms, it becomes the negative
of the other. This suggests altering the contour $C_1$ in the integral over
$z$ to the contour $C_2$, which is depicted in Figure 2 . This enables
us to write the heat kernel in the following form,

\begin{figure}
\epsffile{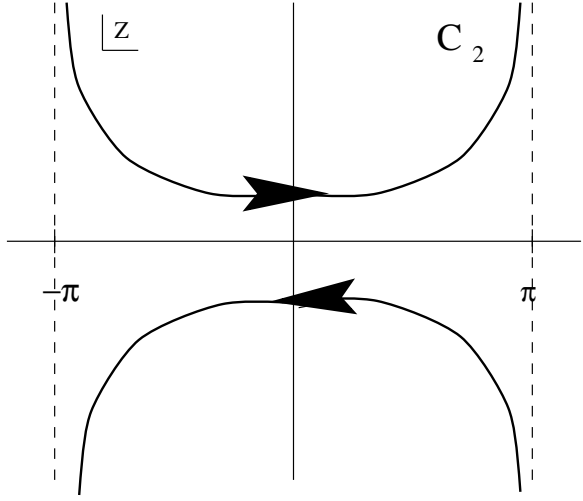}
\caption{Integration contour $C_2$ in the complex z plane}
\end{figure}

\begin{eqnarray}
  F^E(x,x^{\prime },\tau ) = \frac 1{(2\pi )^4\alpha (2\tau )}\sqrt{\frac \pi
\tau }e^{-\frac{(\rho ^2+\rho ^{\prime }{}^2)}{4\tau }-\frac{(z-z^{\prime
})^2}{4\tau }-M^2\tau } \hspace {3cm} \nonumber \\ 
\int_{-\infty }^\infty d\omega  e^{-\tau \omega ^2-i\omega (\overline{t}-
\overline{t}^{\prime }  -\frac S\alpha (\phi -\phi ^{\prime }))}\int_{C_2}
\frac{e^{\frac{\rho \rho ^{\prime }}{2\tau }\cos z-i{\rm Int}(\omega S)\frac{(\phi
-\phi ^{\prime })}\alpha -i\frac z\alpha {\rm Frac}(\omega S)}dz}{e^{-\frac
i\alpha (z-(\phi -\phi ^{\prime }))}-1}.
\end{eqnarray}

\section{Manipulation of the contours}

\paragraph{}
Having recast the problem into one involving contour integration, we
can now manipulate the contour to make evaluation of the
integral easier. In particular, it will be seen that it is easy to pick out
the (rotating, Euclideanized) Minkowski space contribution to the heat kernel
(in other words terms which one would expect in flat, non-conical, but
rotating space). The remainder should be considered as a correction to
the rotating Minkowski space result caused by a combination of the
conical and spinning nature of the space.

\paragraph{}
Examining the integrand of (19) carefully, we see that there are poles
all along the real axis, at $z=(\phi -\phi ^{\prime
})+2\pi m\alpha $. By manipulating the contour $C_2$ into the contour $C_3$,
as shown in Figure 3, it is easy to pick out the contribution of the
$m=0$ pole. The contours in the right hand diagram
which do not encircle the pole at $m=0$ are designated $C_3$. The
curves constituting $C_3$ can always be chosen so they pass
arbitrarily close to the $m=0$ pole, and hence we pick up no other
polar contribution inside our contours of integration. The residue of 
this pole (remembering the integration is clockwise), gives a
contribution to the heat kernel $F_p^E(x,x^{\prime },\tau )$, explicitly

\begin{figure}
\epsffile{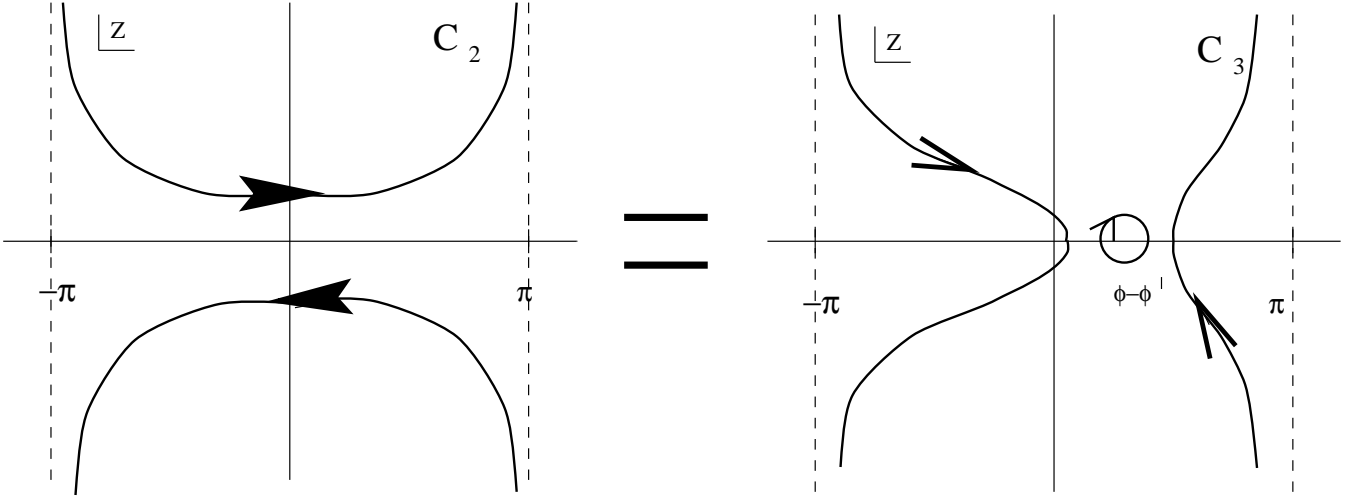}
\caption{Manipulation of the contours in the complex z plane}
\end{figure}

\begin{eqnarray}
F_p^E(x,x^{\prime },\tau ) = \frac{\sqrt{\frac \pi \tau }e^{\frac{\rho \rho ^{\prime }}{2\tau }\cos (\phi -\phi^{\prime })}}{(2\pi )^3(2\tau )
}e^{-\frac{(\rho ^2+\rho ^{\prime }{}^2)}{4\tau }-\frac{(z-z^{\prime })^2}{
4\tau }-M^2\tau } \int_{-\infty }^\infty d\omega e^{-\tau \omega ^2-i\omega (
\overline{t}-\overline{t}^{\prime })}  \nonumber \\
=\frac 1{(4\pi \tau )^2}e^{-\left( \frac{(\overline{t}-\overline{t}
^{\prime })^2+({\bf r}-{\bf r}^{\prime })^2+(z-z^{\prime })^2}{4\tau }
+M^2\tau \right) } \hspace{1cm},
\end{eqnarray}

\noindent where ${\bf r} = (\rho, \phi) $. Thus we can write the total
result for the heat kernel as

\begin{equation}
F^E (x, x^{\prime}, \tau) = F_p^E(x, x^{\prime}, \tau)+ F_{\alpha}^E (x,
x^{\prime}, \tau) \hspace{1cm},
\end{equation}

\noindent where

\begin{eqnarray}
F_{\alpha}^E (x, x^{\prime}, \tau) = \frac{1}{(2 \pi)^4 \alpha (2 \tau)} 
\sqrt{\frac{\pi}{\tau}} e^{- \left( \frac{(\rho-\rho^{\prime})^2
+(z-z^{\prime})^2}{4 \tau} +M^2 \tau \right)} \hspace{3.5cm} \nonumber \\
\int_{-\infty}^{\infty} d \omega e^{-\tau \omega^2 -i \omega (\overline{t}-%
\overline{t}^{\prime}-\frac{S}{\alpha}(\phi-\phi^{\prime}) )} \int_{C_3} 
\frac{ e^{-\frac{\rho \rho^{\prime}}{ \tau} \sin^2 \left( \frac{z}{2}
\right) -i {\rm Int} (\omega S) \frac { (\phi-\phi^{\prime})}{\alpha} -i \frac{z}{%
\alpha} {\rm Frac}(\omega S)}dz}{e^{-\frac{i}{\alpha} (z-(\phi-\phi^{\prime}))} -1}
\hspace{0.25cm}. 
\end{eqnarray}

\section{Evaluating the trace of the heat kernel}

\paragraph{}
The expression for the determinant of the Klein-Gordon operator on
this manifold requires only the trace of the heat kernel (see section
10). Therefore we will henceforth specialize to the case $x=x^{\prime }$
. This simplifies matters considerably, and allows us to write the heat
kernel in the following form,

\begin{eqnarray}
 F^E (x,x, \tau) =\frac{1}{(4 \pi \tau)^2} e^{-M^2 \tau} +  \hspace{7cm} \nonumber \\ \hspace{2cm} \frac{\sqrt{\frac{\pi}{\tau} } e^{-M^2 \tau}}{(2 \pi)^4 \alpha (2 \tau)}
\int_{-\infty}^{\infty} d \omega e^{-\tau \omega^2} \int_{C_3} \frac{e^{-%
\frac{\rho^2}{\tau} \sin^2 \left( \frac{z}{2} \right) - i \frac{z}{\alpha}
{\rm Frac} (\omega S)} dz}{e^{-\frac{i}{\alpha} z} -1} \hspace{0.5cm}.
\end{eqnarray}

\paragraph{}
The first term of this expression is exactly the same as one would obtain in
flat, non-rotating Minkowski space, the rest is a correction due to the
conical and spinning nature of the space. If we make the following definition,

\begin{equation}
I(z,S) = \int_{-\infty}^{\infty} d \omega e^{- \tau \omega^2 -i \frac{z}{%
\alpha} {\rm Frac} (\omega S)} \hspace{1cm},
\end{equation}

\noindent we can write the expression thus,

\begin{equation}
F^E (x, x, \tau) = \frac{e^{-M^2 \tau}}{(4 \pi \tau)^2} \left[ 1+
  \frac{1}{2 \pi \alpha} \sqrt{\frac{\tau}{\pi}} \int_{C_3} \frac{I(z, S) e^{-\frac{%
\rho^2}{\tau} \sin^2 \left(\frac{z}{2} \right)} dz}{e^{-\frac{i}{\alpha}z}-1}
\right] \hspace{1cm}.
\end{equation}

\section{The conical space-time with flux}

\paragraph{}
Our next task will be to integrate over the position variable $x$ in order to
obtain the trace of the heat kernel. After this we will then perform the
contour integral over $z$,  writing the heat kernel in terms of a
single integration variable $\omega $. The procedure used is
somewhat lengthy, so in order to demonstrate the technique it seems
appropriate that we first consider the simpler case of a conical singularity
with non-zero flux (but zero rotation).

\paragraph{}
In the case of non-zero flux, the metric is the same as for an
infinitesimally thin line source at the origin (1), but the novel feature is
that the fields upon this space-time are ``twisted'', the amount of twisting
being dictated by a parameter $\sigma $ (see \cite{mor1}). After finding
the eigenfunctions in this space-time which satisfy the cylindrical symmetry
and twisted field requirements, we can go through exactly the same procedure
as above to derive an expression for the heat kernel.

\paragraph{}
For example, the Green's function for this problem is required to display
the following symmetry,

\begin{equation}
G_F^E (t, \rho, \phi + 2 \pi \alpha, z) = e^{2 \pi i \sigma} G_F^E (t, \rho,
\phi, z) \hspace{1cm}.
\end{equation}

\noindent The eigenfunctions of the Klein-Gordon operator on the
conical manifold satisfying this twisted boundary condition are

\begin{equation}
\Psi_{\omega, k, k_3, m} (x)= \frac{1}{(2 \pi)^{\frac{3}{2}} \alpha^{\frac{1%
}{2}} } J_{\frac{|m+\sigma|}{\alpha}} (k \rho) e^{i \left( k_3 z - \omega t
+ (m+\sigma) \frac{\phi}{\alpha} \right)} \hspace{1cm}.
\end{equation}

\noindent This system bears considerable similarity to the
case we are interested in, but because $\sigma $ is a constant and not a
function of $\omega $, it is somewhat simpler. We can proceed through
analogous steps to those above, to obtain the following expression,

\begin{eqnarray}
F^E(x,x^{\prime },\tau ) =\frac 1{(2\pi )^4\alpha (2\tau )}\sqrt{\frac \pi
\tau }e^{-\frac{(\rho ^2+\rho ^{\prime }{}^2)}{4\tau }-\frac{(z-z^{\prime
})^2}{4\tau }-M^2\tau } \hspace{3.5cm} \nonumber \\ 
\int_{-\infty }^\infty d\omega e^{-\tau \omega ^2-i\omega (t-t^{\prime
})+i\frac \sigma \alpha (\phi -\phi ^{\prime })}\int_{C_2}\frac{e^{\frac{
\rho \rho ^{\prime }}{2\tau }\cos z-i{\rm Int}(\sigma )\frac{(\phi -\phi ^{\prime
})}\alpha -i\frac z\alpha {\rm Frac}(\sigma )}dz}{e^{-\frac i\alpha (z-(\phi -\phi
^{\prime }))}-1} \hspace{0.25cm}.
\end{eqnarray}

\paragraph{}
As before, we can distort the contour to pick out the Minkowski space
contribution, but because $\sigma $ is not a function of $\omega $, we can
now perform the $\omega $ integration as well, which is simply Gaussian.
Thus if we set the variable $x = x^{\prime}$, we obtain the following,

\begin{equation}
F^E (x, x, \tau) = \frac{e^{-M^2 \tau}}{(4 \pi \tau)^2} \left[ 1+ \frac{1}{2
\pi \alpha} \int_{C_3} \frac{ e^{-\frac{\rho^2}{\tau} \sin^2 \left(\frac{z}{2%
} \right)-\frac{iz}{\alpha} {\rm Frac} (\sigma) } dz}{e^{-\frac{i}{\alpha}z}-1}
\right] \hspace{1cm}.
\end{equation}

\paragraph{}
We now take the trace of the heat kernel by integrating over $x$. Naturally
this generates the usual volume divergences in the expression, but any local
divergences are where the interest lies. Let us write $%
\int dzdt=V_2$, and $\int \rho d\rho d\phi =\alpha U_2$ where we
integrate over the (infinite) volume of the space. On this manifold
there is no problem with closed time-like curves, because we have no 
rotation, so the space-time is well defined everywhere. Therefore
after integration we obtain the following expression,

\begin{equation}
Tr(F^E (x, x^{\prime}, \tau)) = \frac{e^{-M^2 \tau}}{(4 \pi \tau)^2} \left[
\alpha V_2 U_2 + \frac{V_2 \tau}{2} \int_{C_3} \frac{e^{- \frac{iz}{\alpha}
{\rm Frac} (\sigma) } dz}{(e^{-\frac{i}{\alpha}z} -1) \sin^2 \left( \frac{z}{2}
\right)} \right] \hspace{0.5cm}.
\end{equation}

\paragraph{}
The only thing left to do is to evaluate the contour integral. It should be
noted that the integrand is well behaved (goes to zero) at positive and
negative imaginary infinity. Therefore we can close the contour $C_3$ at
positive and negative imaginary infinity, thus encircling the only pole in
the integrand, which is at the origin. Therefore we need merely to find the
residue of the pole at the origin.

\paragraph{}
The easiest way to evaluate this is to Taylor expand the denominator for
small $z$. This gives the result

\begin{equation}
(1-\cos z)(e^{-\frac{i}{\alpha} z}-1) = \frac{-i}{2 \alpha} z^3 (1-\frac{iz}{%
2 \alpha} -\frac{z^2}{12} - \frac{z^2}{6 \alpha^2})+{\cal O} (z^6) 
\hspace{0.5cm}.
\end{equation}

\noindent Inverting the denominator, we arrive at the result

\begin{equation}
\frac{1}{(1-\cos z)(e^{-\frac{i}{\alpha} z}-1)} = 2 i \alpha \frac{\left(1+ 
\frac{iz}{2 \alpha} +\frac{z^2}{12} \left(1-\frac{1}{\alpha^2} \right)
\right)}{z^3}+{\cal O} (1) \hspace{0.5cm}.
\end{equation}

\noindent Since only the $z^{-1}$ terms contribute to the integral, it
is necessary to Taylor expand the numerator too. If we call the
numerator $J(z)$, then

\begin{equation}
J(z) = J(0) + zJ^{\prime}(0) + \frac{z^2}{2}
J^{\prime\prime}(0)+{\cal O} (z^3) \hspace{1cm},
\end{equation}

\noindent so we obtain

\begin{eqnarray}
Tr(F^E (x, x^{\prime}, \tau)) = \frac{e^{-M^2 \tau}}{(4 \pi \tau)^2} \Bigg[
\alpha V_2 U_2 + \hspace{6cm} \nonumber \\ \hspace{5cm} 2 i \alpha V_2 \tau \oint \frac{\frac{J(0)}{12} \left(1-%
\frac{1}{\alpha^2} \right) +i \frac{J^{\prime}(0)}{2 \alpha} + \frac{%
J^{\prime\prime}(0)}{2}}{z} \Bigg],
\end{eqnarray}

\noindent and finally

\begin{eqnarray}
Tr(F^E (x, x^{\prime}, \tau)) = \frac{e^{-M^2 \tau}}{(4 \pi \tau)^2} \left[
\alpha V_2 U_2 - 4 \pi \alpha V_2 \tau \left( \frac{J(0)}{12} \left( 1-\frac{%
1}{\alpha^2} \right) \hspace{1cm} \right. \right.  \nonumber \\
\left. \left. \hspace{1cm} +i \frac{J^{\prime}(0)}{2 \alpha} + \frac{J^{\prime\prime}(0)}{%
2} \right) \right] \hspace{0.25cm}.
\end{eqnarray}

\noindent In this case, $J(0) =1$, $J^{\prime}(0) = -\frac{i}{\alpha} {\rm Frac} (\sigma) $
and $J^{\prime\prime}(0) = - \frac{\left({\rm Frac} (\sigma)\right)^2}{\alpha^2} $%
, so we can write

\begin{eqnarray}
Tr \left( F^E (x, x^{\prime}, \tau) \right) = \frac{e^{-M^2 \tau}}{(4 \pi
\tau)^2} \left[ \alpha V_2 U_2 - 4 \pi \alpha V_2 \tau \left( \frac{1}{12}
\left( 1-\frac{1}{\alpha^2} \right) \hspace{1cm} \right. \right.  \nonumber \\
\left. \left. \hspace{1cm} + \frac{{\rm Frac} (\sigma)(1-{\rm Frac}(\sigma))}{2 \alpha^2} \right)
\right] \hspace{0.25cm}.
\end{eqnarray}

\paragraph{}
It should be noted that in addition to the usual Minkowski space volume
divergence $V_2U_2$ (which is here multiplied by a factor of $\alpha $ to
compensate for the volume of a cone compared with globally flat space) there
is an area divergence $V_2$ which is the contribution from the conical
singularity. The standard result for a zero flux conical space-time is
recovered from this result by setting $\sigma =0$. It is interesting to note
that the heat kernel trace depends only on the fractional part of $\sigma $,
something which is alluded to in \cite{mor1}.

\section{Application to the rotating case}

\paragraph{}
If we compare equation (29) with (25), we see that in the rotating space we
have a very similar situation to the non-zero flux case, where $I(z,S)$
plays an analogous role to $J(z)$. There is an additional complication though,
in that we want to restrict our manifold to radii $\rho $ greater than $ 
\overline{\rho}_0+\overline{k}$, to avoid the closed time-like curves and 
potential unitarity problems. Therefore when we take the trace of the
heat kernel, we should be careful to only integrate over the relevant region.

\paragraph{}
Bearing this in mind, it is easy to obtain the following expression for the
trace of the heat kernel,

\begin{eqnarray}
Tr\left( F^E(x,x^{\prime },\tau )\right) =\frac{e^{-M^2\tau }}{(4\pi \tau
)^2} \Bigg[ \alpha V_2(U_2-\pi (\overline{\rho} _0+\overline{k})^2)+ \hspace{3cm} \nonumber \\
\hspace{1.5cm} \frac{\sqrt{\frac \tau \pi }V_2\tau }2\int_{C_3}\frac{I(z,S)e^{-%
\frac{(\overline{\rho} _0+\overline{k})^2}\tau \sin ^2\left( \frac z2\right) }dz}{(e^{-\frac
i\alpha z}-1)\sin ^2\left( \frac z2\right) } \Bigg] \hspace{0.5cm}.
\end{eqnarray}

\paragraph{}
Thus if we identify $J(z)=\sqrt{\frac \tau \pi }I(z,S)e^{-\frac{(\overline{\rho} _0+
\overline{k})^2%
}\tau \sin ^2\left( \frac z2\right) }$ we clearly have an analogous
situation to the case of the conical singularity with non-zero flux. This gives
us the following results,

\begin{eqnarray}
J(0) &=&1 \hspace{1cm}, \\
J^{\prime }(0) &=&\frac{-i}\alpha \sqrt{\frac \tau \pi }\int_{-\infty
}^\infty {\rm Frac}(\omega S)e^{-\tau \omega ^2}  \hspace{1cm}, \\
J^{\prime \prime }(0) &=&\frac{-1}{\alpha ^2}\sqrt{\frac \tau \pi }%
\int_{-\infty }^\infty {\rm Frac}^2(\omega S)e^{-\tau \omega ^2}-\frac{(\overline{\rho}
_0+\overline{k})^2}{2\tau } \hspace{0.5cm}.
\end{eqnarray}

\noindent Consequently we can write the trace of the heat kernel in the following form,

\begin{eqnarray}
Tr \left( F^E (x, x^{\prime}, \tau) \right) = \frac{e^{-M^2 \tau}}{(4 \pi
\tau)^2} \Bigg[ \alpha V_2 U_2 \hspace{6cm}  \nonumber \\ \hspace{5cm}
- 4 \pi \alpha V_2 \tau \left( \frac{1}{12} \left( 1-\frac{1}{\alpha^2%
} \right) + \frac{K(S, \tau)}{2 \alpha^2} \right) \Bigg] \hspace{0.5cm},
\end{eqnarray}

\noindent where

\begin{equation}
K(S, \tau) = \sqrt{\frac{\tau}{\pi}} \int_{-\infty}^{\infty} d \omega
e^{-\tau \omega^2} \left( {\rm Frac}(\omega S) - {\rm Frac}^2 
(\omega S) \right) \hspace{1cm}.
\end{equation}

\paragraph{}
It should be noted that the introduction of a radial boundary at 
$\rho^{\prime} =\rho_0$ to cut out the closed time-like curve region has had 
little impact on expression for the trace of the heat kernel. This is not 
entirely surprising, as it has already been noted \cite{cog1} that the 
mathematics of manifolds with conical singularities bears strong resemblance 
to the mathematics of manifolds with boundaries. In effect, in the simple 
conical singularity case described by metric (1) there is a boundary with 
zero radius, and shifting it to a finite distance away from the origin does
nothing to change the terms which arise purely from its existence.

\paragraph{}
As in the case of the conical singularity with flux, the pure conical
singularity result is regained when we set the rotation parameter $S$ to zero.

\section{Simplifying the result}

\paragraph{}
In order to be able to do calculations with the quantity we have just obtained,
it would be preferable if we could evaluate our expression for $K(S, \tau) $
in a somewhat more palatable form. The key to doing this is noticing that
part of the integrand in $K(S, \tau)$, which we will call $f (\omega, S) $,
is a continuous, even periodic function, with period $\frac{1}{S} $. This
allows us to make a useful expansion of the integrand as a Fourier Series.

\paragraph{}
If we define $K (S, \tau ) = \sqrt{\frac{\tau}{\pi}} \int d \omega f
(\omega, S) e^{- \tau \omega^2} $, we can write $f(\omega, S) $ as

\begin{equation}
f(\omega, S ) = \frac{1}{6} - \sum_{k=1}^{\infty} \frac{\cos (2k \pi S
\omega)}{ (k \pi)^2} \hspace{1cm}.
\end{equation}

\noindent This then allows us to write down a more convenient expression for 
$K(S,\tau) $,

\begin{equation}
K(S, \tau) = \frac{1}{6} - \sum_{k=1}^{\infty} \frac{ e^{-\frac{k^2 \pi^2 S^2%
}{\tau}}}{(k \pi)^2} \hspace{1cm}.
\end{equation}

\paragraph{}
It is obvious from this expression for $K(S,\tau )$ that $K(0,\tau )=0$, and
so our result reproduces the simple conical singularity case when the
rotation parameter $S$ is set to zero. It is also the case that the result
depends only on $S^2$, so there will be no distinction in the
effective action between rotations in opposite senses.

\section{Evaluating the determinant of the Klein-Gordon operator}

\paragraph{}
Having obtained the trace of the heat kernel in a useful form, it is now
fairly straightforward to obtain an expression for the determinant of the
Klein-Gordon operator on this manifold  \cite{fur2} \cite{fur3}
\cite{fur4} \cite{fur1}. Since the eigenvalues of this operator increase
without bound, the value of the determinant itself is
infinite. As is standard, we invoke the following De Witt - Schwinger 
proper time representation for the determinant, which allows us to
isolate (and subsequently remove) this divergence more easily,

\begin{eqnarray}
\log \det (\Box_x+M^2) = Tr \log (\Box_x + M^2) &=& - Tr \int_0^{\infty} \frac{%
d \tau}{\tau} e^{-\tau(\Box_x+M^2)}  \nonumber \\
&=& - \int_0^{\infty} \frac{d \tau}{\tau} Tr e^{-\tau(\Box_x+M^2)}.
\end{eqnarray}

\paragraph{}
It should now be noted that the expression $e^{-\tau(\Box_x+M^2)}$ satisfies
the following differential equation

\begin{equation}
\frac{\partial}{\partial \tau} F^E (x, x^{\prime}, \tau) = -(\Box_x +M^2)
F^E (x, x^{\prime}, \tau) \hspace{1cm},
\end{equation}

\noindent which is simply the heat equation for the differential operator $(\Box_x+M^2)$. Therefore $e^{-\tau (\Box _x+M^2)}$ is a representation of the
heat kernel, and we arrive at the expression

\begin{equation}
\log \det (\Box_x+M^2) = - \int_0^{\infty} \frac{d \tau}{\tau} Tr \left( F^E
(x, x^{\prime}, \tau) \right) \hspace{1cm}.
\end{equation}

\noindent Since we have an expression for $Tr \left( F^E (x,
  x^{\prime}, \tau) \right)$ it is now a simple matter to calculate $\log \det (\Box_x+M^2)$, explicitly

\begin{eqnarray}
- \int_0^{\infty} \frac{d \tau}{\tau} Tr \left( F^E (x, x^{\prime}, \tau)
\right) = - \int_0^{\infty} \frac{d \tau}{\tau} \frac{e^{-M^2 \tau}}{(4 \pi
\tau)^2} \Bigg[ \alpha V_2 U_2 \hspace{3cm}  \nonumber \\ \hspace{4cm}
 - 4 \pi \alpha V_2 \tau \left( \frac{1}{12} \left( 1-\frac{1}{\alpha^2%
} \right) + \frac{K(S, \tau)}{2 \alpha^2} \right) \Bigg].
\end{eqnarray}

\paragraph{}
Some of the integrals in this expression are divergent, so we need to find a
suitable regularization method to make sense of the result. Dimensional
regularization proves useful, where we extend the dimensionality of
space-time to $d$ dimensions. The space-time is considered to have $d-2$ globally flat
dimensions, with the other 2 dimensions forming the truncated cone. In other
words $d=4$ in all above expressions. The expression for $Tr\left(
F(x,x^{\prime },\tau )\right) $ then becomes

\begin{eqnarray}
Tr \left( F_d^E (x,x^{\prime},\tau) \right)= \frac{e^{-M^2 \tau}}{(4 \pi
\tau)^{\frac{d}{2}}} \Bigg[ \alpha V_{d-2} U_2 \hspace{6cm}  \nonumber \\
\hspace{4cm} - 4 \pi \alpha V_{d-2} \tau \left( \frac{1}{12} \left( 1-\frac{1}{%
\alpha^2} \right) + \frac{K(S, \tau)}{2 \alpha^2} \right) \Bigg] \hspace{0.5cm}.
\end{eqnarray}

\paragraph{}
If we then let $d=4-\epsilon $, we can write $V_{d-2}=\mu ^\epsilon V_2$,
where $\mu $ is just an arbitrary parameter with the dimensions of mass, 
included to keep the overall dimensions correct. This allows us to
write

\begin{equation}
Tr \left( F_d^E (x,x^{\prime},\tau) \right)= (4 \pi \tau \mu^2)^{\frac{%
\epsilon}{2}} Tr \left( F_4^E (x, x^{\prime}. \tau) \right) \hspace{1cm}.
\end{equation}

\noindent The intention is now to calculate integral (48) with $F_d^E(x,x^{\prime
},\tau )$ instead of $F_4^E(x,x^{\prime },\tau )$, and then afterwards let $%
\epsilon \rightarrow 0$. Concentrating initially on the first two terms in (48),
we obtain after integration (and some initial regulation),

\begin{eqnarray}
\frac{-\alpha V_2U_2}{16\pi ^2}\left( \frac{4\pi \mu ^2}{M^2}\right)
^{\frac \epsilon 2}M^4\Gamma (-2+\frac \epsilon 2) \nonumber
\hspace{6cm} \\ \hspace{3cm} + \frac{4\pi \alpha V_2}{16\pi ^2}\left( \frac 1{12}\left( 1-\frac 1{\alpha
^2}\right) \right) \left( \frac{4\pi \mu ^2}{M^2}\right) ^{\frac \epsilon
2}M^2\Gamma (-1+\frac \epsilon 2) \hspace{0.5cm}. 
\end{eqnarray}

\paragraph{}
The Gamma function has poles along the negative real axis for integer 
values, which are regulated by the $\epsilon $ factors. The $\epsilon $
poles resulting from this need to be isolated and
included in counter-terms if we intend to consider field theory on the
manifold (see section 14).

\paragraph{}
It is easy to extract the finite part of these terms by using an expansion of
the Gamma function. Explicitly

\begin{eqnarray}
\Gamma (-2 + \frac{\epsilon}{2}) &=& \frac{1}{2} \left[ \frac{2}{\epsilon} + 
\frac{3}{2} - \gamma \right] + {\cal O} (\epsilon) \hspace{1cm}, \\
\Gamma (-1 + \frac{\epsilon}{2}) &=& - \left[ \frac{2}{\epsilon} +1 -\gamma
\right] + {\cal O} (\epsilon) \hspace{1cm},
\end{eqnarray}

\noindent where $\gamma $ is the Euler-Mascheroni constant. The finite
pieces of the first two terms of the logarithm of the determinant of
the Klein-Gordon operator are then

\begin{eqnarray}
-\frac{\alpha V_2 U_2 M^4}{16 \pi^2} \left[ \frac{3}{4} - \frac{\gamma}{2} +%
\frac{1}{2} \log \left( \frac{4 \pi \mu^2}{M^2} \right) \right] \hspace{5cm}\nonumber \\
\hspace{2cm} + \frac{4 \pi \alpha V_2 M^2}{16 \pi^2} \left( \frac{1}{12} \left(1-\frac{1}{%
\alpha^2} \right) \right) \left[ \gamma -1 - \log \left( \frac{4 \pi \mu^2}{%
M^2} \right) \right] \hspace{0.5cm}.
\end{eqnarray}

\paragraph{}
The integration of the final term of (48) is more interesting.
It should first be noted that $K(S,\tau)$ contains a constant
term which is independent of $S$, and also a sum of terms which are
$S$ dependent. Since $K(0,\tau )=0$, this implies that if we consider 
the integration of $K(S,\tau)$ in (48), the resultant (correctly regularized) 
terms should all vanish when $S=0$. In other words, the $S$ dependent 
pieces of the result should still cancel with the $S$ independent term
in the limit $S \rightarrow 0$. However, the integral of the $S$
independent term is formally infinite, for all $S$, for the same
reason as the terms we dealt with above. Since it must cancel with 
the $S$ dependent terms as $ S \rightarrow 0 $, this means that the 
$S$ dependent pieces must also become infinite in this limit. 

\paragraph{}
If we regulate the $S$ independent piece in the above manner to
control its divergence, (which is enough to make the last term of integral (48)
finite for all non-zero $S$), then it will no longer cancel with the
$S$ dependent terms in the limit $S \rightarrow 0 $, since these terms
still diverge. Therefore we must find a way to regulate the $S$
dependent pieces too, so that they will be finite in the $S
\rightarrow 0 $ limit, and properly cancel with the regulated $S$ 
independent piece. Bearing this in mind, and noting the following
integral identity,

\begin{equation}
\int_0^{\infty} x^{\nu-1} e^{-\frac{\beta}{x}} e^{-\gamma x} dx = 2 \left( 
\frac{\beta}{\gamma} \right)^{\frac{\nu}{2}} K_{\nu} (2 \sqrt{\beta \gamma} )
\hspace{1cm},\end{equation}

\noindent where $\beta $ and $\gamma $ are real and positive and $K_\nu (z)$ is a
modified Bessel function of the second kind, we can write the integral
of one of the S dependent terms in the sum as follows,

\begin{equation}
\int_0^{\infty} \frac{d \tau}{\tau^2} e^{-\frac{k^2 \pi^2 S^2}{\tau}}
e^{-M^2 \tau}= \frac{2M}{k \pi S} K_1 (2k \pi SM) \hspace{1cm}.
\end{equation}

\paragraph{}
As expected, this expression diverges in the limit $S \rightarrow 0 $. Consequently
the integral needs to be regulated and we must remove terms in $S$ which
are potentially divergent as $S \rightarrow 0 $. These subtractions will
be absorbed in the redefinitions of the bare coupling constants of
field theories on the manifold. We can be guided by the following
power series expansion,

\begin{equation}
K_1(z)=\frac 1z+\log \left( \frac z2\right) I_1(z)-\frac z4\sum_{m=0}^\infty
\left[ \Psi (m+1)+\Psi (m+2)\right] \frac{\left( \frac{z^2}4\right) ^m}{%
m!(1+m)!} ,
\end{equation}

\noindent where $\Psi (z)$ is the Psi (di-gamma) function.

\paragraph{}
It is tempting to merely subtract all the terms, using the above power
series expansion, which diverge as $S \rightarrow 0$. However,
such subtractions should not spoil the large $S$ behaviour of the $S$ 
dependent terms (they go to zero). We therefore regulate expression
(56) by replacing the right hand side with the following expression

\begin{eqnarray}
M^2 \Upsilon(k, S, M, \delta) = M^2 \left[ \frac{2 K_1 (2k\pi SM)}{k \pi S M} -
  \frac{1}{k^2 \pi^2 S^2 M^2}  \hspace{3cm} \right. \nonumber \\  \hspace{7cm} \left.
-\frac{2 K_1 (2k \pi S \delta)}{k \pi S  \delta} +\frac{1}{k^2 \pi^2 S^2 \delta^2} \right]\hspace{0.25cm},
\end{eqnarray}

\noindent where $\delta $ is another parameter with the dimensions of
mass, in the same fashion as $\mu$. Its exact value will be determined
by renormalization equations in section 14. Instead of diverging in 
the limit $ S \rightarrow 0$, $\Upsilon(k, S, M, m) $
tends to the finite value $ \log (M^2 / \delta^2 ) $. The terms we have
added to the result of (56) in order to achieve this can be regarded
as redefinitions of the coupling constants of field theories on our
manifold (see section 14)

\paragraph{}
The $S$ independent piece also needs regulation, but as noted earlier 
it is merely another Gamma function-like term and can be regulated as 
follows,

\begin{equation}
\int_0^{\infty} \frac{d \tau}{\tau^2} \frac{1}{6} e^{-M^2 \tau} = \frac{M^2}{%
6} \left[ \gamma -1 - \log \left( \frac{4 \pi \mu^2}{M^2} \right)
\right] \hspace{1cm}.
\end{equation}

\noindent This result relates the arbitrary parameter $\mu$ to the
arbitrary parameter $\delta$, since
we require that this expression cancels with the $S$ dependent terms when 
$S \rightarrow 0 $. Hence $ \mu^2 = \frac{\delta^2}{4 \pi} e^{\gamma-1}
$. Therefore we can write the completely regulated integral 
of expression (44) as

\begin{equation}
\int_0^\infty \frac{d\tau }{\tau ^2}K(S,\tau ) =\frac{M^2}6\log
\left(\frac{M^2}{\delta^2} \right) -\sum_{k=1}^\infty \frac{M^2}{k^2\pi
  ^2}  \Upsilon(k,M, \delta , S). 
\end{equation}

\paragraph{}
Combining all our results together, we obtain that the correct, regularized
result for the $\log \det (\Box_x+M^2) $ on our conical, spinning manifold is

\begin{eqnarray}
\log \det (\Box _x+M^2) =\frac{\alpha V_2U_2M^4}{16\pi ^2}\left[-\frac
14+\frac 12 \log \left( \frac{M^2}{\delta^2}\right) \right] \hspace{2cm}
\nonumber \\
+\frac{4\pi \alpha V_2M^2}{16\pi ^2}\left( \frac 1{12}\left( 1-\frac
1{\alpha ^2}\right) \right) \log \left( \frac{ M^2}{
\delta^2}\right) \hspace{1cm} \nonumber \\
\hspace{1.5cm} +\frac{4\pi \alpha V_2 M^2}{16\pi ^2\alpha ^2}\left[ \frac 1{12} \log
\left( \frac{M^2}{\delta^2} \right)-\sum_{k=1}^\infty \frac {M^2}{2k^2\pi ^2}
\Upsilon (k,M,\delta ,S) \right] \hspace{0.25cm}.
\end{eqnarray}

\section{The physics of the space-time}

\paragraph{}
It is well known from study of thermodynamics that a complete description of
the thermal behaviour of a physical system can be obtained if one knows the
partition function for that system. It is also well known that one can
obtain results about quantum systems in which temperature plays no part by
using similar mathematical techniques. This methodology is known as the path
integral formalism of quantum field theory, a useful reference being \cite
{ryd1}. If one considers a self interacting scalar field theory with
potential $V(\phi (x))$, the action for the theory can be written

\begin{equation}
S(\phi) = \int \sqrt{-g} d^4 x \left (\frac{1}{2} \partial_{\mu} \phi
\partial^{\mu} \phi - V(\phi) \right) \hspace{1cm}.
\end{equation}

\paragraph{}
It should be noted that if we wish to use metric (2) as the space-time
background $g_{\mu \nu }$ for our quantum fields, $g=\det g_{\mu \nu }$
changes sign for $\rho ^{\prime }>\frac{|S|}\alpha -k$ and $\rho
^{\prime }<\frac{|S|}\alpha -k$. This means that the Hamiltonian of
a scalar field on this background would not be Hermitian in the region $\rho
^{\prime }<\frac{|S|}\alpha -k$, which is another reason why it was
excluded from the manifold.

\paragraph{}
With Euclideanization, and an integration by parts the above expression
becomes

\begin{equation}
S(\phi )=iS_E(\phi )=i\int \sqrt{g}d^4x\left( \frac 12\phi \Box _x\phi
+V(\phi )\right) \hspace{1cm}.
\end{equation}

\noindent The partition function (or transition amplitude as it is known in field
theory) is then defined as

\begin{equation}
Z=\int D\phi e^{iS(\phi )}=\int D\phi e^{-S_E(\phi )} \hspace{1cm}.
\end{equation}

\paragraph{}
If we assume $V(\phi )$ has a minimum at $\phi (x)=\phi _0$, where $\phi _0$
is a constant (not a function of $x$), we can expand the action around this
minimum point to obtain a first order (saddle point) approximation to the
partition function $Z$. For example, if $V^{\prime }(\phi _0)=0$ and $%
V^{\prime \prime }(\phi _0)=M^2$, we can write

\begin{equation}
S_E(\phi _0+\phi )=S_E(\phi _0)+\frac 12\int \sqrt{g}d^4x\phi \left[ \Box
_x+M^2\right] \phi \hspace{1cm}.
\end{equation}

\noindent This gives as a first order approximation to the partition function

\begin{equation}
Z_1=e^{-X_4V(\phi _0)-\frac \hbar 2\log \det (\Box _x+M^2)} \hspace{1cm},
\end{equation}

\noindent where $X_4$ is the total volume of the space, and we have temporarily
re-inserted the factor of $\hbar $ .

\paragraph{} The saddle-point first order approximation to the effective action is then defined as
the negative of the logarithm of the above partition function, namely

\begin{equation}
W_1(\phi _0)=X_4V(\phi _0)+\frac \hbar 2\log \det (\Box _x+M^2) \hspace{1cm}.
\end{equation}

\noindent A term proportional to classical potential at its minimum value is clearly
included in this effective action, along with a term proportional to $\hbar $%
. Now the reason for concentrating on evaluating the logarithm of the determinant of the
Klein-Gordon operator on our manifold becomes clear - it allows us to
directly calculate the first quantum correction term to the classical action.

\paragraph{}
Considering now the classical field $\phi _c(x)$ \cite{ryd1}, we can expand
perturbatively around its minimum value at $\phi _0$ to obtain the first order 
($\hbar$) approximation to the effective action for classical fields on our 
spinning manifold. We consider a typical self interacting potential, $V(\phi_c
)=\frac{m^2\phi_c^2}2+\frac{\lambda \phi_c^4}{24},$ so consequently, inserting 
$M^2=m^2+\frac{\lambda \phi _c^2}2$ and defining $\bf{r}= (\rho^{\prime}, \phi)$
we obtain as the first order approximation to the effective action 
with Minkowskiian signature

\begin{eqnarray}
&&\Gamma (\phi _c) =\int_{\rho ^{\prime}>\rho _0}\sqrt{-g}d^4x\left[ \frac
12 \partial_{\mu} \phi _c \partial^{\mu} \phi _c-\frac{m^2\phi _c^2}2-\frac{\lambda \phi _c^4}{24}
\right]  \nonumber \\&&
+\int_{\rho ^{\prime}>\rho _0}\sqrt{-g}d^4x(1+\alpha \pi \rho _0^2\delta
^{(2)}({\bf {r}-{r_0}))}\frac{M^4}{32\pi ^2}\left[- \frac 14+\frac
  12 \log \left( \frac{M^2}{\delta^2}\right) \right]  \nonumber \\ &&
\left. +\int_{\rho ^{\prime }>\rho _0}\sqrt{-g}d^4x\delta ^{(2)}({\bf {r}-{
r_0})}\frac{4\pi \alpha M^2}{32\pi ^2}\left( \frac 1{12}\left( 1-\frac
1{\alpha ^2}\right) \right) \left[\log \left(
\frac{M^2}{\delta^2}\right) \right] \right.  \nonumber \\ 
&& \left. +\int_{\rho ^{\prime }>\rho _0}\sqrt{-g}d^4x\delta ^{(2)}({\bf {r}-{
r_0})}\frac{4\pi \alpha M^2}{32\pi ^2\alpha ^2}\left[ \frac 1{12}\log
\left( \frac{M^2}{\delta^2} \right)- \right. \right. \nonumber \\ && 
\hspace{6cm} \left. \left. \sum_{k=1}^\infty \frac 1{2k^2\pi ^2} 
\Upsilon (k,M,\delta,S) \right] \right].
\end{eqnarray}

\paragraph{}
The effective action is expressed entirely in terms of the classical 
field $\phi _c$, coupling constants, physical variables and the (as
yet) undetermined parameter $\delta$ . It has been successfully
regularized, but the coupling constants of the theory still need to 
be renormalized to provide us with a complete
description of the system. This will be considered in section 14. It 
needs to be borne in mind that the minimum $\phi _0$ of the potential 
we have chosen is either zero if $m^2$ is positive, or 
$\phi _0=\sqrt{\frac{-6m^2}\lambda }$ if $m^2$ is considered to be 
negative. Thus we will continue to denote the minimum point simply
by $\phi _0$ to accommodate the possibility of spontaneous symmetry breaking.

\section{The interior metric : an example}

\paragraph{}
So far we have only considered the properties of the exterior metric, i.e.
the one which is supposed to describe the space-time at radii $\rho ^{\prime
}>\rho _0$. There are a variety of models that one could choose for the
interior metric \cite{jen2} \cite{sol2} \cite{sol4}, all varying in terms of
ease of manipulation and realism in representing a supposed physical system.
In order to demonstrate how they can be used, and consequently the complete
manifold be considered, we will look at the simplest case here.

\paragraph{}
The simplest model to consider is that of a infinitely thin rotating shell of matter
at radius $\rho ^{\prime }=\rho _0$. The metric inside this shell is then
accurately described by the metric of a rotating disk, namely

\begin{eqnarray}
ds^2 = -\left( \sqrt{1+\frac{S^2 (\rho_0^2 - {\rho^{\prime}}^2)}{\rho_0^4}}
dt + \frac{S \left( \frac{\rho^{\prime}}{\rho_0} \right)^2 d \phi}{\alpha 
\sqrt{1+\frac{S^2 (\rho_0^2 - {\rho^{\prime}}^2)}{\rho_0^4}} } \right)^2 +
dz^{\prime}{}^2 +d {\rho^{\prime}}^2  \nonumber \\
+ \left( \frac{{\rho^{\prime}}^2}{\alpha^2} + \frac{S^2 \left( \frac{%
\rho^{\prime}}{\rho_0} \right)^4}{\alpha^2 (1+ \frac{S^2 (\rho_0^2 - {%
\rho^{\prime}}^2)}{\rho_0^4} )} \right) d \phi^2 \hspace{1cm},
\end{eqnarray}

\noindent where $(t, \rho^{\prime}, \phi, z^{\prime}) = (t, \rho^{\prime}, \phi+2 \pi \alpha,
z^{\prime}) $ , $0 < \rho^{\prime}< \rho_0$, $0 < \phi < 2 \pi \alpha $, $-
\infty< t < \infty$, $-\infty < z^{\prime} < \infty$ and  $0< \alpha < 1$.

\paragraph{}
Although the above metric has terms containing the constant $\alpha $, it
should be noted that there is no deficit angle in this interior space-time.
It should also be noted that when $\rho ^{\prime }=\rho _0$ this metric
becomes

\begin{equation}
ds^2= -(dt + \frac{S}{\alpha} d \phi )^2 + dz^{\prime}{}^2 + d {\rho^{\prime}}^2 +
\left( \frac{\rho_0^2+S^2}{\alpha^2} \right) d \phi^2 \hspace{1cm}.
\end{equation}

\paragraph{}
Since we require the metric components to patch up at $\rho ^{\prime }=\rho
_0$, this determines for us the value of the constant $k$ in equation (2)
for this interior model, explicitly $k=-\rho _0+\frac{\sqrt{\rho _0^2+S^2}}%
\alpha $.

\paragraph{}
If we now examine the condition on $\rho _0$ from section 1 which prevents
the occurrence of closed time-like curves in the space-time , $\left( \rho
_0>\frac{|S|}\alpha -k\right) $, we see that this choice of interior
space-time and hence constant $k$ satisfies the inequality. Thus we can 
be assured that causality, unitarity and such things are all intact on the 
complete manifold.

\paragraph{}
Despite looking rather complicated, the metric (69) is indeed flat, as can
be verified by making the change of variables $d \phi = d \phi^{\prime}+ 
\frac{ \alpha S}{\rho_0^2} dt $. The metric then becomes

\begin{equation}
ds^2 = -\left( 1+\frac{S^2}{\rho_0^2} \right) dt^2 +\frac{{\rho^{\prime}}^2}{%
\alpha^2} d {\phi^{\prime}}^2 +d {\rho^{\prime}}^2 +dz^{\prime}{}^2 \hspace{1cm}.
\end{equation}

\paragraph{}
Since $0 < \phi^{\prime}< 2 \pi \alpha $ it is clear that the original
metric (69) is merely a flat disc viewed from a rotating co-ordinate system.
This fact allows us to place an upper bound on the variable $S$. We cannot 
consider a rotating cylinder which has superluminal velocity, so this 
constrains $ S< \rho_0 $.

\paragraph{}
It is possible to re-write this metric with a Euclidean signature by using
the same transformations as in section 2, $\rho^{\prime}=i \rho $ 
$z^{\prime} =iz $. We then obtain

\begin{equation}
ds^2 = -\left( 1+\frac{S^2}{\rho_0^2} \right) d t^2 -\frac{{\rho}^2}{%
\alpha^2} d {\phi^{\prime}}^2 -d \rho^2 -dz^2 \hspace{1cm}.
\end{equation}

\paragraph{}
Although both the interior and exterior space-times can be seen to be
locally flat, this does not imply that there is no curvature at all in the
complete space-time - curvature may still exist at the boundary between the two
space-times, $\rho ^{\prime }=\rho _0$. Although the interior and exterior
metric coefficients have the same value at the boundary, the derivatives of
those coefficients are not the same there. This means that the extrinsic
curvature of the boundary is different for the two sides, and thus the net
curvature is the difference between the interior and exterior extrinsic
curvatures at the boundary.

\paragraph{}
The energy-momentum tensor of the singular boundary surface was calculated
by this method in \cite{jen2}. The non-zero components of the four
dimensional energy-momentum tensor describing the cylinder at $\rho^{\prime}
 = \rho_0$ can be written (using metric (69)) as

\begin{eqnarray}
8 \pi G T_0^0 &=& \frac{2 \pi \alpha \rho_0}{\sqrt{\rho_0^2 +S^2}} \left(
\alpha-\frac{(\rho_0^2+S^2)^{\frac{3}{2}}}{\rho_0^3} \right) \delta^{(2)} (%
{\bf {r}-{r_0})}  \nonumber \\
8 \pi G T_1^1 &=& \frac{2 \pi \alpha \rho_0}{\sqrt{\rho_0^2 +S^2}} 
\left(\frac{S^2 (\rho_0^2+S^2)^{\frac{1}{2}}}{\rho_0^3} \right) 
\delta^{(2)} ({\bf {r}-{%
r_0})}  \nonumber \\
8 \pi G T_2^2 &=& \frac{2 \pi \alpha \rho_0}{\sqrt{\rho_0^2 +S^2}} \left(
\alpha-\frac{(\rho_0^2+S^2)^{\frac{1}{2}}}{\rho_0} \right) \delta^{(2)} (%
{\bf {r}-{r_0})}  \nonumber \\
8 \pi G T_1^0 =-8 \pi G T_0^1 &=& \frac{2 \pi \alpha \rho_0}{\sqrt{\rho_0^2 +S^2}} \left(\frac{%
S (\rho_0^2+S^2)}{\rho_0^3} \right) \delta^{(2)} ({\bf {r}-{r_0})} 
\hspace{1cm},
\end{eqnarray}

\noindent where ${\bf {r}=(\rho^{\prime}, \theta)}$ and $0 < \theta <
2 \pi \alpha$. The curvature is easily obtained if we note the
Einstein equation for general relativity and take the trace, i.e. 
$R = -Tr (8 \pi G T_{\mu}^{\nu}). $ Hence

\begin{equation}
R(\alpha ,S)=\frac{4\pi \alpha }{\sqrt{\rho _0^2+S^2}}\left( \sqrt{\rho
_0^2+S^2}-\alpha \rho _0\right) \delta ^{(2)}({\bf {r}-{r_0})} \hspace{1cm}.
\end{equation}

\paragraph{}
Expressions for the mass and angular momentum of the cylinder can be simply
obtained from (73). To contrast with the infinitesimally thin conical singularity case (1), 
it can be seen that the expression for the source mass will involve both the spin 
parameter $S$ and the deficit angle parameter $\alpha$. What this means is 
that for a cylinder with given physical mass and angular momentum, we must compute 
$\alpha$ and $S$ using expressions for the physical quantities derived from the 
energy momentum tensor. This system also allows the possibility of a cylinder with
non-zero mass and angular momentum, but with the deficit parameter $\alpha$ =1. This
appears to be a model of a domain wall, which separates two topologically 
distinct regions of Minkowski space.

\paragraph{}
It should also be noted that when $S=0$ and $\rho _0=0$ we return to the
expression for curvature for the simple conical singularity obtained in \cite
{ger2}, \cite{man1} and other papers.

\section{Heat kernel in the interior region}

\paragraph{}
In order to complete the calculation of the effective action on the whole
space-time, we need to consider the expression for the heat kernel in the
interior region. Fortunately this is fairly straightforward, since the
interior region is flat and has no deficit angle. If we consider the metric

\begin{equation}
ds^2=-{dt_i}^2-\rho^2 d\phi ^2-dz^2-d{\rho}^2 \hspace{1cm},
\end{equation}

\noindent where $t_i=\sqrt{1+\frac{S^2}{\rho _0^2}} t$ , and 
$0<\rho < \overline{\rho} _0 $, this is clearly a special
case of metric (1), where we have changed the variables and let $\alpha
=1 $. We would therefore expect the heat kernel to be written

\begin{equation}
F^E(x,x,\tau )=\frac{e^{-M^2\tau }}{(4\pi \tau )^2}\left[ 1+\frac 1{2\pi
}\int_{C_3}\frac{e^{-\frac{\rho^2}\tau \sin ^2\left( \frac z2\right) }dz%
}{e^{-iz}-1}\right] \hspace{1cm}.
\end{equation}

\noindent Taking the trace of this equation gives

\begin{equation}
Tr \left(F^E(x,x^{\prime },\tau )\right)=\frac{e^{-M^2\tau }\sqrt{1+\frac{S^2}{\rho _0^2}}%
V_2}{(4\pi \tau )^2}\left[ \pi \overline{\rho} _0^2 +\frac \tau
2\int_{C_3}\frac{1-e^{-\frac{{\overline{\rho} _0}^2}{\tau } \sin ^2\left( \frac
z2\right) }dz}{\sin ^2\left( \frac z2\right) (e^{-iz}-1)}\right].
\end{equation}

\paragraph{}
Inspection of this result and reviewing section 7 convinces us that this is
zero. Therefore the only internal contribution to the effective action in this model 
comes from the classical terms

\begin{equation}
\Gamma (\phi _c)=\int_{\rho ^{\prime }<\rho _0}\sqrt{-g}d^4x\left[ \frac
12 \partial_{\mu} \phi _c \partial^{\mu} \phi _c-\frac{m^2\phi _c^2}2-
\frac{\lambda \phi _c^4}{24} \right] \hspace{1cm}. 
\end{equation}

\section{Renormalization of the coupling constants}

\paragraph{}
Now that we have an expression for the effective action on the complete
manifold, we can discuss the renormalization of the coupling constants which
occur in the theory. The approach used here is based on the method discussed
in \cite{cog1}. All the coupling constants and parameters of the bare
Lagrangian need renormalization since we are dealing with an interacting
field theory. The typical terms which arise in the Lagrangian for a self
interacting scalar field on curved background (including a term for the
gravitational Lagrangian itself) are \cite{bir1} 

\begin{equation}
{\mathcal L}  (\phi _{c,B},R)
=\frac 12\phi _{c,B} \Box _x\phi _{c,B}+\frac{m_B^2}2\phi
_{c,B}^2 + \frac{\lambda_B \phi _{c,B}^4}{4!} +
\varepsilon_B R + \frac 12 \xi_B R\phi _{c,B}^2 ,
\end{equation}

\noindent where the subscript B denotes a
bare parameter. The first part of the renormalization procedure is to
redefine the bare parameters so that they remove the divergent pieces
in the effective action, calculated in section 10. These may be
written

\begin{eqnarray}
 -\frac{\alpha V_2 U_2}{16 \pi^2}  M^4 \left(\frac{1}{\epsilon}
  \right) + 
\frac{4\pi \alpha V_2}{16\pi ^2} \frac {M^2}{12} \left( 1-\frac 1{\alpha
^2} \right) \left(\frac{-2}{\epsilon} 
  \right) \hspace{3.5cm} \nonumber \\ + \frac{4 \pi \alpha V_2}{16
\pi^2 \alpha^2} \left[ \frac{M^2}{12} \left( \frac{-2}{\epsilon} \right)-
\sum_{k=1}^{\infty} \frac{1}{2k^2 \pi^2} \left(\frac{1}{k^2 \pi^2 S^2}
\right. \right.  \hspace{4cm} \nonumber \\  \left. \left.  \hspace {4cm} + \frac{2 M^2 K_1 (2k\pi S \delta)}{k\pi S \delta} -\frac{M^2}{k^2 \pi^2 S^2 \delta^2}
  \right)  \right] \hspace{0.5cm}, \end{eqnarray}

\noindent where as before $M^2=m^2 + \frac{\lambda}{2} \phi_c^2 $. By
redefining the (physically meaningless) bare parameters to cancel
these terms, we can obtain the finite effective action derived in
section 10. Explicitly, by comparing powers of $\phi_c$ in the
expressions, we are lead to the following results, where we have
written $\frac{1}{2} \xi_B R = \sigma_B \delta^{(2)} ({\bf r-r_0})$
and $ \varepsilon_B R = \nu_B \delta^{(2)} ({\bf r-r_0})$,

\begin{eqnarray}
\phi_{c,B} &=& \mu^{-\frac{\epsilon}{2}} \phi_c \\
m_B^2 &=& m^2 \left( 1 + \frac{\hbar \lambda}{16 \pi^2 \epsilon} \right) + 
{\mathcal O} (\hbar^2) \\
\lambda_B &=&  \mu^{\epsilon} \left( \lambda + \frac{3 \lambda^2}{16
    \pi^2} \frac{\hbar}{\epsilon}  \right)+{\mathcal O} (\hbar^2) \\
\nu_B &=& \mu^{-\epsilon} \left( \nu + \frac{4\pi \alpha m^2}{16
    \pi^2} \left( \frac{1}{12} \left( 1-\frac{1}{\alpha^2} \right) \right)
\frac{\hbar}{\epsilon} + \frac{4 \pi \alpha}{16 \pi^2}
  \frac{m^2}{12} \frac{\hbar}{\epsilon} + {\mathcal O} (\hbar^2) \right. \nonumber \\ &&
\left.
+\frac{4\pi \alpha \hbar}{32 \pi^2 \alpha^2}  
\sum_{k=1}^{\infty} \frac{1}{2k^2
\pi^2} \left[ \frac{1}{k^2 \pi^2 S^2} (1-\frac{m^2}{\delta^2}) +\frac{2m^2 K_1 (2k \pi 
  S \delta)}{k \pi S \delta } \right] \right)  \\
\sigma_B &=& \sigma  + \frac{4\pi \alpha }{16
    \pi^2} \left( \frac{1}{12} \left( 1-\frac{1}{\alpha^2} \right) \right)
\frac{\hbar}{\epsilon} \frac{\lambda}{2} + \frac{4 \pi \alpha}{16 \pi^2} 
\frac{1}{12} \frac{\hbar}{\epsilon} \frac{\lambda}{2} \nonumber \\ &&
+\frac{4\pi \alpha \hbar}{32 \pi^2 \alpha^2}  
\sum_{k=1}^{\infty} \frac{1}{2 k^2
\pi^2} \left[ \frac{\lambda K_1 (2k \pi 
  S\delta)}{2k \pi S \delta} - \frac{\lambda}{2k^2 \pi^2 \delta^2 S^2} \right]
 + {\mathcal O} (\hbar^2) \hspace{0.15cm}. 
\end{eqnarray}

\paragraph{}
It should be noted that because of the particular regularization
scheme that we have chosen to employ, the structure of the
ultra-violet poles in the theory when spin is included has been 
somewhat obscured. We will address these issues in the final part of 
this paper, section 15. However, it is the finite part of the
effective action that is the more important piece, certainly as far 
as physical applications are concerned, and it is correctly given 
by expression (60). 

\paragraph{}
Although we have now successfully removed all divergent pieces of
the effective action, additional redefinitions of the coupling constants
are required to ensure that the physical parameters (e.g. the mass m) 
have the same values (obtained by differentiating the Lagrangian)
after the quantum corrections have been included. To achieve this we
must add further counterterms to the Lagrangian.

\paragraph{}
We can now work with a completely finite effective potential for the
our theory, inferred from the expressions for the effective action in
both the interior and exterior regions and expressed in terms of the
renormalized couplings $m$, $\lambda$, etc. , namely

\begin{eqnarray}
V(\phi _c,\alpha ,S) =\left[ \frac{m^2\phi _c^2}2+\frac{\lambda \phi _c^4}{
24}\right] +\hspace{7cm} \nonumber \\
\frac{M^4}{32\pi ^2}\left[ -\frac 14+\frac 12 \log
\left( \frac{M^2}{\delta^2}\right) \right] \left( 1-\frac{\pi \rho _0^2}
\alpha \left( \sqrt{1+\frac{S^2}{\rho _0^2}}-\alpha ^2\right) \delta ^{(2)}(
{\bf {r}-{r_0})}\right) \nonumber \\
 +\frac{4\pi \alpha M^2}{32\pi ^2}\left( \frac 1{12}\left( 1-\frac
1{\alpha ^2}\right) \right) \log \left( \frac{ M^2}{
\delta^2}\right) \delta ^{(2)}({\bf {r}-{r_0})} +\hspace{2.5cm} \nonumber \\
  \hspace{0.25cm} \frac{4\pi \alpha }{32\pi ^2\alpha ^2}\left[ \frac{M^2}{12}\log
\left( \frac{M^2}{\delta ^2} \right)-\sum_{k=1}^\infty 
\frac {M^2}{k^2\pi ^2} \left[ \left( \frac{K_1(2
k\pi SM)}{ k\pi SM}-\frac{1}{2 k^2 \pi ^2 S^2 M^2}\right)
\right.  \right. \nonumber \\
\left.  \left. -\left( \frac{K_1(2k\pi \delta S)}{k\pi \delta S}-\frac{1}{2k^2\pi
^2 \delta^2 S^2}\right) \right] \right] \delta ^{(2)}({\bf {r}-{r_0})} \hspace{0.5cm} . 
\end{eqnarray}

\noindent It will be more helpful to rewrite this expression in terms of the
space-time curvature $R$,

\begin{eqnarray}
V(\phi _c,\alpha ,S) =\left[ \frac{m^2\phi _c^2}2+\frac{\lambda \phi _c^4}{
24}\right] + \hspace{7cm} \nonumber \\
\frac{M^4}{32\pi ^2}\left[ -\frac 14+\frac 12 \log \left( 
\frac{M^2}{\delta^2}\right) \right] \left( 1-\frac{\rho _0\sqrt{\rho
_0^2+S^2}}{4\alpha ^2}\frac{\left( \sqrt{1+\frac{S^2}{\rho _0^2}}-\alpha
^2\right) }{\left( \sqrt{1+\frac{S^2}{\rho _0^2}}-\alpha \right) }R\right) 
\nonumber \\
\left. +\frac{M^2}{32\pi ^2}\frac{\sqrt{\rho _0^2+S^2}}{\sqrt{\rho _0^2+S^2
}-\alpha \rho _0}\left( \frac 1{12}\left( 1-\frac 1{\alpha ^2}\right)
\right) \log \left( \frac{M^2}{\delta^2}\right)
R\right.  \nonumber \hspace{1.5cm}\\ 
\left. +\frac{M^2}{32\pi ^2\alpha ^2}\frac{\sqrt{\rho _0^2+S^2}}{\sqrt{
\rho _0^2+S^2}-\alpha \rho _0}\left[ \frac 1{12}\log \left( \frac{M^2}{\delta
^2} \right)-\sum_{k=1}^\infty \frac 1{k^2\pi ^2} \hspace{3cm} \right. \right.  \nonumber \\
\left. \left( \frac{K_1(2 k\pi SM)}{ k\pi SM}-\frac 1{2
^2k^2\pi ^2S^2M^2}\right) -\left( \frac{K_1(2k\pi \delta S)}{k\pi
\delta S}-\frac{1}{2k^2\pi ^2 \delta^2 S^2}\right) \right] R.
\end{eqnarray}

\paragraph{}
We now impose the following renormalization condition,

\begin{equation}
m^2=\left. \frac{\partial ^2V}{\partial \phi _c^2}\right| _{\phi
  _c=0,R=0} \hspace{1cm}. 
\end{equation}

\noindent With some algebra we obtain the expression 
\begin{equation}
\frac{m^2\lambda ( -\log \left( \frac{ \delta ^2}{m^2}\right) )}{%
32\pi ^2}=0 \hspace{1cm}.
\end{equation}

\noindent This result enables us to fix the parameter $\delta $ in our
previous expressions, giving $\delta ^2=m^2$. We are then left with the 
effective potential

\begin{eqnarray}
V(\phi _c,\alpha ,S) =\left[ \frac{m^2\phi _c^2}2+\frac{\lambda \phi _c^4}{
24}\right]+ \hspace{7cm}  \nonumber \\
\frac{M^4}{64\pi ^2}\left[ -\frac 12+\log \left( \frac{M^2}{m^2}\right)
\right] \left( 1-\frac{\rho _0\sqrt{\rho _0^2+S^2}}{4\alpha ^2}\frac{\left( 
\sqrt{1+\frac{S^2}{\rho _0^2}}-\alpha ^2\right) }{\left( \sqrt{1+\frac{S^2}{
\rho _0^2}}-\alpha \right) }R\right) \hspace{1cm} \nonumber \\
\left. +\frac{M^2}{32\pi ^2}\frac{\sqrt{\rho _0^2+S^2}}{\sqrt{\rho _0^2+S^2
}-\alpha \rho _0}\frac 1{12}\left( 1-\frac 1{\alpha ^2}\right) \log \left( 
\frac{M^2}{m^2}\right) R\right.  \hspace{3.5cm} \nonumber \\
\left. +\frac{M^2}{32\pi ^2\alpha ^2}\frac{\sqrt{\rho _0^2+S^2}}{\sqrt{
\rho _0^2+S^2}-\alpha \rho _0}\left[ \frac 1{12}\log \left( \frac{M^2}{m^2}
\right) -\sum_{k=1}^\infty \frac 1{k^2\pi ^2} \hspace{3cm} \right. \right.  \nonumber \\
\left. \left( \frac{K_1(2k\pi S M)}{k\pi SM}-\frac{1}{
2k^2\pi ^2S^2M^2}\right) -\left( \frac{K_1(2k\pi Sm)}{k\pi Sm}-\frac
1{2k^2\pi ^2S^2m^2}\right) \right] R.
\end{eqnarray}

\paragraph{}
Although we have now fixed all the arbitrary parameters in the
effective potential, we have yet to impose several renormalization
conditions on the other coupling constants in the theory. We satisfy
these conditions by adding further counterterms, effectively re-defining
the coupling constants once again, although this time not by an infinite 
amount. The next condition we impose is

\begin{equation}
\left. \frac{\partial ^4V}{\partial \phi _c^4}\right| _{\phi _c=\phi
_{0,}R=0}=\lambda \hspace{1cm}. 
\end{equation}

\noindent This gives the counter-term

\begin{equation}
\frac{\lambda ^2}{64\pi ^2}\left( \frac{8m^4}{M_0^4}+\frac{8m^2}{M_0^2}%
-22-6\log \left( \frac{M_0^2}{m^2}\right) \right) \frac{\phi _c^4}{4!}
\hspace{1cm}, 
\end{equation}

\noindent where $M_0^2=m^2+\frac \lambda 2\phi _0^2$. We can view this
term as arising from a redefinition of $\lambda$. 

\paragraph{}
We must now look at the
counter-terms which arise from terms proportional to $R.$ This is more
complicated than in the non-spinning case, because $R$ depends on two
variables here, not one. However when $R$ is small the two variables $%
(\alpha ,S)$ decouple from one another ($\delta R=\left. \frac{\partial R}{%
\partial \alpha }\right| _{\alpha =1,S=0}d\alpha +\left. \frac{\partial R}{%
\partial S}\right| _{\alpha =1,S=0}dS+...$), and we can treat the two
limiting cases ($\alpha =1,S=0$) separately. 

\paragraph{}
The next renormalization condition can be (somewhat symbolically)
written as 

\begin{equation}
\left. \frac{\partial V}{\partial R}\right| _{\phi_c =0,R=0}=0 \hspace{1cm}.
\end{equation}

\noindent When $\phi_c =0 $ then $M=m$, so we merely need consider the
expression $V(0, \alpha, S)$, 

\begin{equation}
V(0, \alpha, S) = -\frac{m^4}{64\pi ^2}\left( \frac 12
\right) \left( 1-\frac{\rho _0\sqrt{\rho _0^2+S^2}}{4\alpha ^2}\frac{\left( 
\sqrt{1+\frac{S^2}{\rho _0^2}}-\alpha ^2\right) }{\left( \sqrt{1+\frac{S^2}{
\rho _0^2}}-\alpha \right) }R\right) \hspace{0.25cm}.
\end{equation}

\paragraph{}
If we set $\alpha =1$ and differentiate, we obtain

\begin{equation}
\left. \frac{\partial V(0,1,S)}{\partial (R(1,S))}\right| _{S=0} =
\frac{m^4}{64\pi ^2}\left( \frac 12
\right) \frac{\rho _0^2}4 \hspace{1cm}.
\end{equation}

\noindent Considering now the other variable, set $S=0$ and
differentiate, 

\begin{equation}
\left. \frac{\partial V(0,\alpha,0)}{\partial (R(\alpha,0))}\right| _{\alpha=1} =
\frac{m^4}{64\pi ^2}\left( \frac 12
\right) \frac{2\rho _0^2}4 \hspace{1cm}.
\end{equation}

\noindent Thus we derive the counterterm

\begin{equation}
-\frac{m^4}{64\pi ^2}\frac 12\frac{\rho _0^2}4(R(1,S)+2R(\alpha ,0)) \hspace{1cm}.
\end{equation}

\paragraph{}
The next renormalization condition to consider is 

\begin{equation}
\left. \frac{\partial ^3V%
}{\partial R\partial \phi ^2}\right| _{\phi =\phi _{0,}R=0}=0 \hspace{1cm}.
\end{equation}
 
\noindent Once again we regard the $S=0$ and $\alpha=1$ cases
separately, and in the former case we obtain the counterterm

\begin{eqnarray}
\left[ \frac \lambda {64\pi ^2}\left( 4(M_0^2-m^2)-(6M_0^2-4m^2)\log \left( 
\frac{m^2}{M_0^2}\right) \right) \frac{2\rho _0^2}4+\right. \nonumber \\ 
\left.\frac \lambda {64\pi ^2}
\frac{\left( 3M_0^2-2m^2+M_0^2\log \left( \frac{M_0^2}{m^2}\right) \right) }{
3M_0^2}\right] \frac 12R(\alpha ,0)\phi _c^2 \hspace{1cm}.
\end{eqnarray}

The $\alpha=1$ case is slightly more complicated, because the corresponding 
expression diverges as $S \rightarrow 0 $. Therefore instead of
subtracting a constant part of $\xi$, we must subtract an $S$ dependent piece from $\xi$
in order to remove this divergence and satisfy (98).  This leads us to the 
following expression for the counterterm, 

\begin{eqnarray}
\frac{\rho_0^2}{4} \frac{\lambda}{64 \pi^2} \left(
  4(M_0^2-m^2)+(6M_0^2-4m^2) 
\log \left( \frac{M_0^2}{m^2} \right) \right) \frac{1}{2} R(1,S)
\phi_c^2  - \nonumber \\
\frac{\rho_0^2}{32 \pi^2 S^2} \frac{\partial^2}{\partial \phi^2} \left[ \frac{M^2}{6} \log \left( \frac{M^2}{m^2} \right) - 
\sum_{k=1}^\infty \frac {M^2}{k^2\pi ^2} \Upsilon (k,M,\delta,S) \right]_{\phi=\phi_0} 
\frac{1}{2} R(1,S) \phi_c^2.  
\end{eqnarray}
\noindent The term in the square brackets in the previous expression
has the form
\begin{eqnarray}
C(S,M_0,m,\lambda) = \sum_{k=1}^{\infty} \frac{\lambda}{k^2 \pi^2} 
\left[ \log \left( \frac{M_0^2}{m^2} \right) + \frac{2 \lambda}{M_0^2}
  (M_0^2-m^2)  
+ \right. \nonumber \\ \left.  2 K_0 (2k \pi S M_0)
+1 -4 \lambda \frac{k \pi S (M_0^2-m^2)}{M_0} K_1 (2k \pi S M_0)
  \right. \nonumber \\ \left.  - \frac{2K_1 (2k \pi S m)}{k \pi S m} 
+\frac{1}{k^2 \pi^2 S^2 m^2} \right] \hspace{0.5cm}.\end{eqnarray}
\noindent The counterterms can then be included in the effective action as follows,

\begin{eqnarray}
&{}& V(\phi _c,\alpha ,S) =\left[ \frac{m^2\phi _c^2}2+\frac{\lambda \phi _c^4}{
24}\right] \nonumber \\ 
&+&\frac{\lambda ^2}{64\pi ^2}\left( \frac{8m^4}{M_0^4}+\frac{8m^2}{
M_0^2}-22-6\log \left( \frac{M_0^2}{m^2}\right) \right) \frac{\phi _c^4}{4!}
  \nonumber \\
&+& \frac{M^4}{64\pi ^2} \left[ -\frac 12+\log \left( \frac{M^2}{m^2}\right)
\right] \left( 1-\frac{\rho _0\sqrt{\rho _0^2+S^2}}{4\alpha ^2}\frac{\left( 
\sqrt{1+\frac{S^2}{\rho _0^2}}-\alpha ^2\right) }{\left( \sqrt{1+\frac{S^2}{
\rho _0^2}}-\alpha \right) }R \right) \nonumber \\
&-& \frac{M^2}{32\pi ^2}\frac{\sqrt{\rho _0^2+S^2}}{\sqrt{\rho _0^2+S^2
}-\alpha \rho _0}\frac 1{12}\left( 1-\frac 1{\alpha ^2}\right) \log \left( 
\frac{m^2}{M^2}\right) R   \nonumber \\
&+& \frac{M^2}{32\pi ^2\alpha ^2}\frac{\sqrt{\rho _0^2+S^2}}{\sqrt{
\rho _0^2+S^2}-\alpha \rho _0}\left[ \frac 1{12}\log \left( \frac{M^2}{m^2}
\right) -\sum_{k=1}^\infty \frac {M^2}{k^2\pi ^2}  \right. \nonumber \\
&{}& \left.  \left( \frac{ K_1(2k\pi SM)}{k\pi SM}-\frac{1}{
2k^2\pi ^2S^2 M^2}\right) -\left(\frac{K_1(2k\pi Sm)}{k\pi Sm}-\frac
1{2k^2\pi ^2S^2m^2}\right) \right] R \nonumber \\
&-&\frac{m^4}{64\pi ^2}\frac 12\frac{\rho _0^2}4(R(1,S)+2R(\alpha ,0)) \hspace{7cm} \nonumber \\
&+&\left[ \frac \lambda {64\pi ^2}\left( 4M_0^2-4m^2+(6M_0^2-4m^2)\log \left( 
\frac{M_0^2}{m^2}\right) \right) \frac{2\rho _0^2}4+ \hspace{2cm}
\right. \nonumber \\ &{}& \left. \hspace{2cm} \frac \lambda {64\pi ^2} \frac{\left( 3M_0^2-2m^2+M_0^2\log 
\left( \frac{M_0^2}{m^2}\right) \right) }{3M_0^2}\right] 
\frac 12R(\alpha ,0)\phi _c^2  \nonumber \\ 
&+& \frac{\rho_0^2}{4} \frac{\lambda}{64 \pi^2} \left(
  4(M_0^2-m^2)+(6M_0^2-4m^2) \log \left( \frac{M_0^2}{m^2} \right)
\right) \frac{1}{2} R(1,S) \phi_c^2 \nonumber \\ &-&
\frac{\rho_0^2}{32 \pi^2 S^2} C(S,M_0,m,\lambda) \frac{1}{2} R(1,S) \phi_c^2 \hspace{0.5cm}.  
\end{eqnarray}

\paragraph{}
We have now renormalized all the parameters present in the bare free
Lagrangian and also the coupling $\lambda $. The final counter-term is
generated by the constraint $V(\phi _{0,}1,0)=0$ and can be written

\begin{eqnarray}
-\frac{m^2\phi_0^2}2-\frac{\lambda \phi_0^4}{24}-\frac{\lambda ^2}{64\pi ^2}
\left( \frac{8m^4}{M_0^4}+\frac{8m^2}{M_0^2}-22-6\log \left( \frac{M_0^2}{m^2}
\right) \right) \frac{\phi_0^4}{4!} + \nonumber \\ 
\frac{M_0^4}{64\pi ^2} \left[ \frac
12-\log \left( \frac{M_0^2}{m^2} \right) \right] \hspace{1cm}. 
\end{eqnarray}

\paragraph{}
Having now calculated all the necessary counterterms, we can write the final expression for the effective potential as follows,

\begin{eqnarray}
&{}& V(\phi _c,\alpha ,S) =\left[ \frac{m^2\phi _c^2}2+\frac{\lambda \phi _c^4}{
24}\right] -\left[ \frac{m^2\phi _0^2}2-\frac{\lambda \phi _0^4}{24}\right] 
\nonumber \\&+&
\frac{\lambda ^2}{64\pi ^2}\left( \frac{8m^4}{M_0^4}+\frac{8m^2}{M_0^2}
-22-6\log \left( \frac{M_0^2}{m^2}\right) \right) \left[ \frac{\phi
_c^4-\phi _0^4}{4!}\right]  \nonumber \\
&+&\frac{M^4}{64\pi ^2}\left[ \log \left( \frac{M^2}{m^2}\right) -\frac
12\right] \left( 1-\left( \frac{\pi \rho _0^2}\alpha \sqrt{1+\frac{S^2}{\rho
_0^2}}-\alpha ^2\right) \delta ^{(2)}({\bf {r}-{r_0})}\right) \nonumber \\
&+& \frac{M_0^4}{
64\pi ^2}\left[ \frac 12-\log \left( \frac{M_0^2}{m^2}\right) \right] 
 +\frac{4\pi \alpha M^2}{32\pi ^2}\frac 1{12}\left( 1-\frac 1{\alpha
^2}\right) \log \left( \frac{M^2}{m^2}\right) \delta ^{(2)}({\bf {r}-{r_0})}
 \nonumber \\
  &+& \frac{4\pi \alpha M^2}{32\pi ^2\alpha ^2}\left[ \frac 1{12}\log
\left( \frac{M^2}{m^2}\right) -\sum_{k=1}^\infty \frac {M^2}{k^2\pi ^2} 
\left( \frac{K_1(2k\pi SM^2)}{k\pi SM^2}-\frac{1}{
2k^2\pi ^2S^2M^2}\right) \right.\nonumber \\ &{}& \hspace{4cm} \left. -\left( \frac{K_1(2k\pi Sm)}{k\pi Sm}-\frac
1{2k^2\pi ^2S^2m^2}\right) \right] \delta ^{(2)}({\bf {r}-{r_0})} \nonumber \\
&-&\frac{4\pi m^4}{64\pi ^2}\frac 12\frac{\rho _0^2}4\left( \frac{\sqrt{\rho
_0^2+S^2}-\rho _0}{\sqrt{\rho _0^2+S^2}}+2\alpha (1-\alpha )\right) \delta
^{(2)}({\bf {r}-{r_0})} \nonumber \\
&+&\left[ \frac \lambda {64\pi ^2}\left( 4M_0^2-4m^2+(6M_0^2-4m^2)\log
\left( \frac{M_0^2}{m^2}\right) \right) \frac{2\rho _0^2}4+\right. \nonumber \\
&{}& \left. \frac \lambda
{64\pi ^2}\frac{\left( 3M_0^2-2m^2+M_0^2\log \left( \frac{M_0^2}{m^2}\right)
\right) }{3M_0^2}\right] \frac 124\pi \alpha (1-\alpha )\phi _c^2\delta ^{(2)}({\bf {r}-{r_0}%
)} \nonumber \\&+& \frac{\rho_0^2}{4} \frac{\lambda}{64 \pi^2} \left( 4(M_0^2-m^2) + \right. \nonumber \\&{}& \left. (6M_0^2-4m^2) \log \left( \frac{M_0^2}{m^2} \right) \right)  2 \pi \left(1-\frac{\rho_0}{\sqrt{\rho_0^2+S^2}} \right) \phi_c^2 \delta^{(2)}({\bf {r}-{r_0})} \nonumber \\&-& \frac{\rho_0^2}{32 \pi^2 S^2} C(S,M_0,m,\lambda)2 \pi \left(1-\frac{\rho_0}{\sqrt{\rho_0^2+S^2}} \right) \phi_c^2 \delta^{(2)}({\bf {r}-{r_0})} \hspace{1cm}.
\end{eqnarray}

\section{Conclusions and discussion}

\paragraph{}
In the course of this article we have employed a variety of
mathematical techniques in order to obtain the renormalized 
first order effective action for massive scalar field theory with 
a self interacting $\lambda \phi ^4$ type potential on the space-time
generated by a rotating cylindrical shell. We have demonstrated
that this theory is well defined, and is not troubled by the closed
time-like curves and unitarity problems that have hindered previous attempts
at quantization on such a space-time. Having obtained the effective 
potential $ V(\phi_c, \alpha, S) $, it can then be used to model
quantum physics on the manifold, which behaves as if we were dealing 
with a classical field theory with field variable $ \phi_c(x) $ and 
potential $ V(\phi_c,\alpha, S) $. It remains to be seen whether there
are any interesting phenomenological consequences of the new terms in
the effective action, particularly with regard to the physics of
cosmic strings and astrophysics in general, and also in
electromagnetic theory, where there is an analogous system in the 
Aharanov-Bohm effect \cite{jen1}.

\paragraph{}
Since many of the novel features in our effective action 
first arose in section 10, where we dealt with the regularization 
and renormalization of the theory, it seems appropriate to expand 
a little on the discussion of the procedure which was used there to 
regulate the $S$ dependent terms in the effective action. A superior 
and more illuminating way in which this regulation can be achieved 
will now be suggested.

\paragraph{}
It will be recalled that in section 10 dimensional regularization was
used to regulate the integral of the $S$ independent term of $K(S, \tau)$
(expression (59)) and then terms were subtracted from the $S$ dependent
parts of $K(S, \tau)$ so that they would cancel with the regulated $S$
independent term in the limit $S \rightarrow 0$ . I would like to suggest 
an alternative regularization procedure, in which integral (59) is
rendered ultra-violet finite in the following manner,

\begin{equation}
\int_0^{\infty} \frac{d \tau}{\tau^2} \frac{1}{6} e^{-M^2 \tau} e^{-
\frac{\eta^2}{\tau}}=\frac{1}{6} \frac{2M}{\eta} K_1 (2 \eta M) \hspace{0.5cm},
\end{equation}

\noindent  $\eta$ being a regulation parameter in the same manner as
$\epsilon$, which was employed previously in section 10. The added 
exponential term in the integrand has the effect of supressing the 
$\tau \rightarrow 0$ (small time, high frequency) contribution to the 
integral and hence provides an ultra-violet cut off, making the 
expression finite for non-zero $\eta$. The $S$ dependent terms of 
$K(S,\tau)$ may also be regulated in this way (they are potentially 
divergent at S=0) and this gives the result

\begin{equation}
\int_0^{\infty} \frac{d \tau}{\tau^2} e^{-M^2 \tau} e^{-\frac{k^2
\pi^2 S^2}{\tau}} e^{-\frac{\eta^2}{\tau}} = 2M \frac{K_1 (2M
\sqrt{\eta^2+k^2 \pi^2 S^2})}{\sqrt{\eta^2+k^2 \pi^2 S^2}} \hspace{0.5cm}.
\end{equation}

\noindent The equivalent of expression (60), although still including
the regulation parameter $\eta$, can therefore be written

\begin{equation}
\frac{1}{6} \frac{2M}{\eta} K_1 (2 \eta M)-2M \sum_{k=1}^{\infty}
\frac{1}{(k\pi)^2}  \frac{K_1 (2M
\sqrt{\eta^2+k^2 \pi^2 S^2})}{\sqrt{\eta^2+k^2 \pi^2 S^2}} \hspace{0.5cm}.
\end{equation}

\paragraph{}
It is now conventional to redefine the bare coupling constants of the
theory to absorb the poles in $\eta$ of the above expression 
as $ \eta \rightarrow 0 $. To that end I introduce the following
counterterms, 

\begin{eqnarray}
\frac{M^2}{6} \left[-\frac{1}{\eta^2 M^2}-2 \frac{K_1(2\eta \delta)}{\eta
    \delta}+\frac{1}{\eta^2 \delta^2} \right] 
 -\sum_{k=1}^{\infty} \frac{M^2}{(k \pi)^2} \left[
    -\frac{1}{M^2(\eta^2+k^2 \pi^2 S^2)} \right. \nonumber \\ \left.
-2 \frac{K_1(2\delta
    \sqrt{\eta^2+k^2\pi^2S^2})}{\delta
    \sqrt{\eta^2+k^2\pi^2S^2}}+\frac{1}{\delta^2(\eta^2+k^2 \pi^2 S^2)}
    \right] \hspace{0.5cm}.
\end{eqnarray}

\noindent The first three terms subtract from expression
(105), whilst the second three terms subtract from (106). It
should be noted that if $S=0$, then all the above terms cancel precisely
with one another. This implies that in the non-spinning limit ($S=0$)
of our spinning space-time, there are no poles in the theory that need 
removing above and beyond those found in physics on the simple conical
singularity manifold described by metric (1). However, for all $S \ne
0 $ the terms in the previous expression no longer all cancel, and 
there are additional ultra-violet poles in the theory, which must 
be removed.

\paragraph{} The nature of these poles is easily deduced by examining the
equations appearing earlier in this article. In mode expansion (8), for
example, the mode frequency $\omega$ is seen to couple to the spin $S$
in the combination $\omega S$. Since quantum theory involves an
integration over all possible mode frequencies, this means that for
{\sl any} $S$ which is non-zero, a sufficiently high frequency mode will
exist which makes $\omega S$ of a significant size. Examining equation
(42), which describes the integration over the mode frequencies, it
can be seen that because part of the integrand is periodic, and does
not fall off at high frequencies, the integrand is bounded by a gaussian
envelope, which does not go to zero fast enough at small $\tau$ to 
supress the energetic modes and make the contribution to the effective
action finite. Additional supression (and hence regularization) is 
required to obtain a finite answer. The physical origin of the
required additional suppression is also easy to understand. Whilst it
is reasonable to insist that the low frequency modes should couple
with the spin like $\omega S$, any spinning object has an angular
frequency associated with its rotation, and one would not expect modes
which have frequency much larger than the angular frequency to be
greatly affected by the object's rotation. In other words, the
interaction between the spin and the modes should fall off as the mode
frequency increases, and this is precisely what the regularization 
procedure accomplishes. The simple boundary condition which allowed us 
to include spin in the conical space-time in the classical theory is 
therefore seen to be too powerful a constraint for the high frequency 
modes in the quantum theory, resulting in the appearance of additional
ultra-violet divergences.  

\paragraph{}
Combining expressions (105) and (106) with terms (108), and setting
$\eta = 0$, we then obtain the finite piece of the integral of 
$K(S, \tau)/\tau^2$, which agrees with the result quoted in (60). This type
of regulation improves upon that used in section 10, because it is clear
from the argument presented here that there are no additional poles
beyond those in the standard non spinning theory as $S \rightarrow 0$. 
This is not readily apparent from expressions (84) and
(85). Effectively in section 10 we neglected the occurence 
of $\eta^2$ in the denominators of several terms in (84) and (85), because
for $S \ne 0$ $\eta^2$ is negligible compared to $k^2 \pi^2 S^2$. 
However if $S=0$ then the $\eta^2$ factors come into play, and 
prevent the occurence of $S$ poles in the theory, replacing them 
with $\eta$ poles instead. These eta poles then cancel with the pole 
resulting from the S independent term, demonstrating that there are no
additional divergences in the $S=0$ theory.

\paragraph{}
In conclusion, we note that the use of the more physical cylindrical
shell source has successfully resolved most of the difficulties that 
have been traditionally associated with this space-time. Models 
involving closed time-like curves always seem to contain a non-physical
component, be it constituent matter which does not satisfy the energy 
conditions, or sources with unphysical amounts of angular
momentum. Once a more physically realistic system is considered,
quantization proves possible, and new ultra-violet divergences are 
discovered, associated with the spin, which do not occur in the simple
conical singularity case. These divergences are related to the
interaction between the high energy modes of quantum fields
propogating on the space-time and the source spin, but it
proves possible to regulate them and renormalize the coupling constants of
the theory so that a consistent description of the theory in all
regimes of $S$ is possible.

\section{Acknowledgments}

\paragraph{}
The author would like to thank E.S. Moreira (Jnr) for alerting him to this
problem in the first place. Thanks also go to D.A. Rasheed and R.S. Farr for
useful discussions, A.J. Gill for helpful suggestions and to
M.B. Green and I.T. Drummond for guidance. Additional inspiration came
from A.C. Faul.


\begin{thebibliography}{99}
\bibitem{aba1}  Abramowitz M. and Stegun I. A. {\sl Handbook of Mathematical
Functions} Dover (1970)



\bibitem{eco1}  Audretsch J. , Economou A and Tsoubelis D {\sl Pair Creation
and Decay of a massive particle near and far away from a Cosmic String} Phys
Rev {\bf D45} (1992) 1103



\bibitem{eco2}  Audretsch J. and Economou A {\sl Quantum Field Theoretical
Processes near Cosmic Strings: Transition Probabilities and Localization}
Phys. Rev. {\bf D44} (1991) 980



\bibitem{bir1}  Birrell N.D. and Davies P.C.W. {\sl Quantum fields in curved
space} CUP (1982)


\bibitem{bec1}  Beckenstein J.D. {\sl Chiral cosmic strings} Phys. Rev. {\bf %
D45} (1992) 2794



\bibitem{cho1}  Cho Y. M. and Park D. H. {\sl Causally spinning anyonic
cosmic string} Phys. Rev. {\bf D46} (1992) R1219



\bibitem{cog1}  Cognola G, Kirsten K and Vanzo L, {\sl Free and
self-interacting scalare fields in the presence of a conical singularity}
Phys. Rev. {\bf D49} (1994) 1029



\bibitem{des1}  Deser S and Jackiw R {\sl Classical and Quantum Scattering
on a Cone} Commun. Math. Phys. {\bf 118} (1988) 495



\bibitem{des2}  Deser S, Jackiw R and 't Hooft G. {\sl Three- Dimensional
Einstein Gravity : Dynamics of Flat Space} Ann. Phys. {\bf 152} (1984) 220



\bibitem{dow1}  Dowker J.S. {\sl Quantum field theory on a cone} J. Phys. A 
{\bf 10} (1977) 115



\bibitem{dow2}  Dowker J.S. {\sl Field Theory around a Cosmic string} Class.
Quant. Grav. {\bf 4} (1987) 157



\bibitem{dow3}  Dowker J.S. {\sl Casimir effect around a cone} Phys. Rev. 
{\bf D36} (1987) 3095



\bibitem{dow4}  Dowker J.S. {\sl Vacuum Averages for Arbitrary Spin around a
cosmic string} Phys. Rev. {\bf D 36} (1987) 3742



\bibitem{dow5}  Dowker J.S. {\sl Heat Kernel Expansion on a Generalized Cone}
J. Math. Phys. {\bf 30} (1989) 770



\bibitem{dow6}  Dowker J.S. {\sl Quantum Field Theory around Conical Defects}
Nuffield Workshop (1989) 251



\bibitem{dow7}  Dowker J.S. {\sl Heat Kernels on curved cones} Class. Quant.
Grav. {\bf 11} (1994) 137



\bibitem{fro1}  Frolov V. P. and Serebryanyi E. M. {\sl Vacuum polarization
in the gravitational field of a cosmic string} Phys. Rev. {\bf D35} (1987)
3779



\bibitem{fur2}  Fursaev D. V. {\sl Local and Global quantum effects on
cosmic string space-time} JETP Lett. {\bf 58} (1993) 479



\bibitem{fur3}  Fursaev D. V. {\sl Spectral Geometry and One Loop
Divergences on manifolds with conical singularities} Phys. Lett. {\bf B334}
(1994) 53



\bibitem{fur4}  Fursaev D. V. {\sl The Heat Kernel expansion on a cone and
quantum fields near Cosmic Strings} Class. Quant. Grav. {\bf 11} (1994) 1431



\bibitem{fur1}  Fursaev D. V. and Solodukhin S. N. {\sl On the description
of the Riemannian Geometry in the presence of conical defects} Phys. Rev. 
{\bf D52} (1995) 2133



\bibitem{gal1}  Gal'tsov D. V. and Letelier P. S. {\sl Spinning Strings and
cosmic dislocations} Phys. Rev. {\bf 47} (1993) 4273



\bibitem{ger1}  de Sousa Gerbert P. and Jackiw R. {\sl Classical and Quantum
Scattering on a Spinning Cone} Commun. Math. Phys. {\bf 124} (1989) 229


\bibitem{ger2}  de Sousa Gerbert P. {\sl On spin and (quantum) gravity in
2+1 dimensions} Nucl. Phys. {\bf B346} (1990) 440



\bibitem{gib1}  Gibbons G.W., Ruiz Ruiz F and Vachaspati T {\sl The
Nonrelativistic Coulomb Problem on a cone} Commun. Math. Phys. {\bf 127}
(1990) 295



\bibitem{gra1}  Gradshteyn I.S. and Ryzhik I.M. {\sl Table of Integrals,
Series, and Products} Academic Press (1980)



\bibitem{gro1}  Gron O. and Soleng H. H. {\sl Gravitational effect of the
quantum vacuum outside a cosmic string} Class. Quant. Grav. {\bf 9} (1992)
1231



\bibitem{har1}  Harari D. D. and Polychronakos A. P. {\sl Gravitational time
delay due to a spinning string} Phys. Rev. {\bf D38} (1988) 3320



\bibitem{haw1}  Hawking, Stephen {\sl Zeta Function Regularization of Path
Integrals in Curved Spacetime} Comm. Math. Phys. {\bf 55} (1977) 133



\bibitem{hel1}  Helliwell T.M. and Konkowski D.A. {\sl Vacuum fluctuations
outside Cosmic Strings} Phys. Rev. {\bf D34} (1986) 1918



\bibitem{hoo1}  't Hooft G. {\sl Nonperturbative two particle scattering
amplitudes in (2+1) dimensional quantum gravity} Commun. Math. Phys. {\bf 117%
} (1988) 685



\bibitem{jen1}  Jensen B. and Kucera J. {\sl On a gravitational
Aharanov-Bohm effect} J. Math. Phys. {\bf 34} (1993) 4975



\bibitem{jen2}  Jensen B. and Soleng H. H. {\sl General relativistic model
of a spinning cosmic string} Phys. Rev {\bf D45} (1992) 3528



\bibitem{jen3}  Jensen B. {\sl Notes on spinning strings} Class. Quant.
Grav. {\bf 9} (1992) L7



\bibitem{kib1}  Kibble T.W.B. {\sl Topology of cosmic domains and strings}
J. Phys {\bf A9} (1976) 1387



\bibitem{let1}  Letelier P.S. {\sl Spacetime defects : torsion loops} Class.
Quant. Grav. (1995) {\bf 12} 2221



\bibitem{let3}  Letelier P.S. {\sl Spinning cosmic strings as torsion line
spacetime defects} Class. Quant. Grav. {\bf 12} (1995) 471



\bibitem{let2}  Letelier P.S. {\sl Spinning strings as torsion line
spacetime defects} Class. Quant. Grav. {\bf 12} (1995) 471



\bibitem{lin1}  Linet B. {\sl Euclidean thermal spinor Green's functions in the
spacetime of a straight cosmic string} Class. Quant. Grav. {\bf 13}
(1996) 1797



\bibitem{lin2}  Linet B. {\sl Euclidean scalar and spinor Green's functions
in Rindler space} e-Print Archive gr-qc/9505033



\bibitem{lin3}  Linet B. {\sl Euclidean spinor Green's functions in the
spacetime of a straight cosmic string} J.Math.Phys. {\bf 36} (1995) 3694



\bibitem{lin4}  Linet B. {\sl The Twisted Euclidean Green's function in the
spacetime of a cosmic string} Int. J. Mod. Phys. {\bf D1} (1992) 371



\bibitem{lin5}  Linet B. {\sl The Euclidean thermal Green function in the
spacetime of a cosmic string} Class. Quantum Grav. {\bf 9} (1992) 2429



\bibitem{lin6}  Linet B. {\sl Quantum Field Theory in the spacetime of a
cosmic string} Phys. Rev. {\bf D35} (1987) 536


\bibitem{lor1} De Lorenci V.A. , De Paula R. and Svaiter N.F. {\sl
    Gravitational particle production in spinning cosmic string
    space-times} e-Print Archive gr-qc/9705049


\bibitem{man1}  Mann R.B. and Solodukhin S. N. {\sl Conical geometry and
quantum entropy of a charged Kerr black hole} Phys. Rev. {\bf \ D54} (1996)
3932



\bibitem{mat1}  Matsas G.E.A. {\sl Semiclassical Gravitational Effects in
the spacetime of a rotating cosmic string} Phys. Rev. {\bf D42} (1990) 2927



\bibitem{maz1}  Mazur P.O. {\sl Spinning Cosmic Strings and Quantization of
Energy} Phys. Rev. Lett. {\bf 57} (1986) 929



\bibitem{maz2}  Mazur P.O. {\sl Induced angular momentum on superconducting
cosmic strings} Phys. Rev. Lett. {\bf D34} (1986) 1925



\bibitem{men1}  Menotti P. and Seminara D. {\sl Closed timelike curves and
the energy condition in 2+1 dimensional gravity} Phys. Lett. {\bf B301}
(1993) 25



\bibitem{mor1}  Moreira (Jnr) E.S. {\sl Massive quantum fields in a conical
background} Nucl. Phys. {\bf B451} (1995) 365



\bibitem{pun1}  Puntigam R.A. and Soleng H.H. {\sl Volterra distortions,
spinning strings and cosmic defects} Class. Quant. Grav. {\bf 14}
(1997) 1129



\bibitem{rus1}  Russell I. H. and Toms D. J. {\sl Symmetry breaking around
cosmic strings} Class. Quantum Grav. {\bf 6} (1989) 1343



\bibitem{ryd1}  Ryder L.H. {\sl Quantum Field Theory }Cambridge University
Press



\bibitem{seb2}  Serebryanyi E. M. {\sl Vacuum Polarization by Magnetic Flux :
The Aharonov Bohm effect} Theor. Math. Phys. {\bf 64} (1985) 846



\bibitem{ser1}  Serebryanyi E.M. , Frolov V.P. and Skarzhinskii {\sl On the
electrodynamical effects in the gravitational field of a Cosmic String}
Moscow 1987, Proceedings, Quantum Gravity 830



\bibitem{kiy1}  Shiraishi K and Hirenzaki S {\sl Quantum aspects of
self-interacting fields around cosmic strings} Class. Quantum Grav. {\bf 9}
(1992) 2277



\bibitem{ska1}  Skarzhinsky V. D. and Harari D. D. {\sl Pair Production in
the Gravitational Field of a Cosmic String} Phys. Lett. {\bf B240} (1990) 322



\bibitem{ska2}  Skarzhinsky V. D. , Harari D. D. and Jasper U. {\sl Quantum
electrodynamics in the gravitational field of a cosmic string} Phys. Rev. 
{\bf D49} (1994) 755



\bibitem{sol1}  Soleng H. H. {\sl Spin Polarized cylinder in Einstein-Cartan
theory} Class. Quant. Grav. {\bf 7} (1990) 999



\bibitem{sol2}  Soleng H. H. {\sl Negative energy densities in extended
sources generating closed timelike curves in General Relativity with and
without torsion} Phys. Rev. {\bf D49} (1994) 1124



\bibitem{sol4}  Soleng H. H. {\sl On the possibility of closed timelike
curves produced by spinning strings in general relativity with and without
torsion} NORDITA-93-62-A



\bibitem{sol3}  Soleng H. H. {\sl A spinning string} Gen. Rel. Grav. {\bf 24}
(1992) 111



\bibitem{sou1}  Souradeep T and Sahni V {\sl Quantum Effects near a point
mass in (2+1) dimensional gravity} Phys. Rev. {\bf D46} (1992) 1616



\bibitem{tod1}  Tod K. P. {\sl Conical Singularities and torsion} Class.
Quant. Grav. {\bf 11} (1994) 1331



\bibitem{wal1}  Wald R. M. {\sl General Relativity} University of Chicago
press

\end{thebibliography}
\end{document}